\documentclass[twocolumn, tighten, times]{aastex631}

\usepackage{graphicx}
\usepackage{epstopdf}
\usepackage{bm}
\usepackage{color}
\usepackage{soul}
\usepackage{threeparttable}  
\usepackage{microtype}
\usepackage{booktabs}
\usepackage{afterpage}

\shorttitle{An updated efficient galaxy morphology classification model}
\shortauthors{Fang. ET AL}
\graphicspath{{./}{figures/}}
\begin{document}
\title{An updated efficient galaxy morphology classification model based on ConvNeXt encoding with UMAP dimensionality reduction}
\author[0000-0001-9694-2171]{Guanwen Fang}
\altaffiliation{Corresponding author: Guanwen Fang}
\affil{School of Mathematics and Physics, Anqing Normal University, Anqing 246011, China; 
\url{wen@mail.ustc.edu.cn}} 
\affil{Institute of Astronomy and Astrophysics, Anqing Normal University, Anqing 246133, China}

\author[0009-0004-0966-6439]{Shiwei Zhu}
\affil{School of Mathematics and Physics, Anqing Normal University, Anqing 246011, China; 
\url{wen@mail.ustc.edu.cn}} 
\affil{Institute of Astronomy and Astrophysics, Anqing Normal University, Anqing 246133, China}

\author[0000-0003-1697-6801]{Jun Xu}
\affil{School of Mathematics and Physics, Anqing Normal University, Anqing 246011, China;
\url{wen@mail.ustc.edu.cn}} 
\affil{Institute of Astronomy and Astrophysics, Anqing Normal University, Anqing 246133, China}

\author[0000-0001-5988-2202]{Shiying Lu}
\affil{School of Mathematics and Physics, Anqing Normal University, Anqing 246011, China;
\url{wen@mail.ustc.edu.cn}} 
\affil{Institute of Astronomy and Astrophysics, Anqing Normal University, Anqing 246133, China}
\affil{Key Laboratory of Modern Astronomy and Astrophysics (Nanjing University), Ministry of Education, Nanjing 210093, China}

\author[0000-0002-5133-2668]{Chichun Zhou}
\affil{School of Engineering, Dali University, Dali 671003, China}

\author[0000-0002-4638-0235]{Yao Dai}
\affil{Shanghai Astronomical Observatory, Chinese Academy of Sciences, 80 Nandan Road, Shanghai 200030, China}
\affil{School of Astronomy and Space Science, University of Chinese Academy of Sciences, No. 19A Yuquan Road, Beijing 100049, China}

\author[0000-0001-8078-3428]{Zesen Lin}
\affil{Institute for Astrophysics, School of Physics, Zhengzhou University, Zhengzhou, 450001, China}

\author[0000-0002-7660-2273]{Xu Kong}
\affil{Department of Astronomy, University of Science and Technology of China, Hefei 230026, China; \url{xkong@ustc.edu.cn}} 
\affil{School of Astronomy and Space Science, University of Science and Technology of China, Hefei 230026, China}
\affil{Institute of Deep Space Sciences, Deep Space Exploration Laboratory, Hefei 230026, China}

\begin{abstract}  
We present an enhanced unsupervised machine learning (UML) module within our previous \texttt{USmorph} classification framework featuring two components: (1) hierarchical feature extraction via a pre-trained ConvNeXt convolutional neural network (CNN) with transfer learning, and (2) nonlinear manifold learning using Uniform Manifold Approximation and Projection (UMAP) for topology-aware dimensionality reduction. This dual-stage design enables efficient knowledge transfer from large-scale visual datasets while preserving morphological pattern geometry through UMAP's neighborhood preservation. We apply the upgraded UML on I-band images of 99,806 COSMOS galaxies at redshift $0.2<z<1.2$ (to ensure rest-frame optical morphology) with $I_{\mathrm{mag}}<25$. The predefined cluster number is optimized to 20 (reduced from 50 in the original framework), achieving significant computational savings. The 20 algorithmically identified clusters are merged into five physical morphology types. About 51\% of galaxies (50,056) were successfully classified. 
To assess classification effectiveness, we tested morphological parameters for massive galaxies with $M_{*}>10^{9}~M_{\odot}$. Our classification results align well with galaxy evolution theory. This improved algorithm significantly enhances galaxy morphology classification efficiency, making it suitable for large-scale sky surveys such as those planned with the China Space Station Telescope (CSST).

\end{abstract}

\section{Introduction} \label{sec:1}
The morphology and structure observed in galaxy images are capable of revealing numerous physical properties, including color, gas content, star formation rate, stellar mass, and environment (e.g., \citealt{kauffmann+2004, omand+2014, kawinwanichakij+2017, Siudek+2022, Huertas-Company+2024, Ghosh+2024, su+2025}). With the rapid development of astronomical observation technologies and the implementation of numerous large-scale sky surveys, an enormous amount of astronomical data has been generated(e.g.,\citealt{Stoughton+2002, Scoville+2007}). Consequently, conventional expert visual inspection (\citealt{Hubble+1926,flugge+1959}) methods have become inadequate for processing such data. Machine learning (ML), owing to its robust data processing capabilities, has emerged as a critical tool in astronomy.

Machine learning (ML) is commonly categorized into two main types: supervised machine learning (SML) and unsupervised machine learning (UML). 
The former one (i.e., SML), such as convolutional neural networks (CNNs; \citealt{Lecun+1998}), through pretraining, can learn hierarchical features from images, enabling detailed classification of galactic fine-scale structures such as bulges, disks, spiral arms, and bars (e.g., \citealt{simmons+2014, Davis+2014, Dieleman+2015, Fernando+2024}). For instance, \cite{Huertas-Company+2015} trained a five-layer CNN using visual classification data from approximately 8,000 H-band galaxies in the CANDELS fields. However, SML requires training with large annotated datasets, and constructing such high-quality annotated datasets often incurs substantial time and resource consumption.

Recent studies have shown that transfer learning and domain adaptation can partly alleviate this requirement by adapting pretrained models to new surveys with only a modest number of labelled targets (e.g., \citealt{DominguezSanchez+2019, Ciprijanovic+2023}). Nevertheless, these methods still rely on some labelled data, and their performance is sensitive to domain shifts such as Point Spread Function (PSF), depth, and redshift, which limits their general applicability. As a result, the latter one (i.e., UML) is widely employed to analyze large-scale unlabeled datasets for classification and annotation. UML can not only extract features from raw images but also cluster galaxies in terms of similar features((e.g., \citealt{Cheng2+2021, Kolesnikov+2023}). For example, \cite{Cheng2+2021} employed a Vector-Quantized Variational Autoencoder (VQ-VAE) combined with hierarchical clustering algorithms to extract features and classify galaxy images from the Sloan Digital Sky Survey (SDSS; \citealt{York+2000}), successfully categorizing galaxies into 27 distinct groups characterized by varying morphological features and physical properties. 
The Self-supervised learning (SSL; \citealt{Gui+2024}) is generally regarded as a branch of UML, as it learns feature representations from unlabeled data through automatically generated supervision signals rather than manual annotations. For instance, the vision transformer (ViT; \citealt{dosovitskiy+2021}) effectively extracts features from similar data domains. Meanwhile, contrastive learning frameworks (e.g., \citealt{Chen+2020}; \citealt{Grill+2020}) learn feature representations by constructing positive and negative sample pairs, obviating the need for explicit data labels.

However, the effectiveness of UML methods strongly depends on the quality of the learned feature representations. In high-dimensional fields, such as image data, the structural complexity of features is pronounced, accompanied by significant noise and redundant information. 
This phenomenon complicates the ability of unsupervised models to identify key feature dimensions, leading to notable degradation in tasks such as feature extraction and cluster analysis (e.g., \citealt{Zhou+2022,Jang+2024}).  
To address the computational and performance limitations of UML in high-dimensional data, dimensionality reduction techniques, such as principal component analysis (PCA; \citealt{Andrzej+1993}), t-distributed stochastic neighbor embedding (t-SNE; \citealt{van+2008}), and uniform manifold approximation and projection (UMAP; \citealt{McInnes+2018}), have been increasingly integrated into UML frameworks. For example, PCA identifies principal components with maximal variance through linear transformation, discarding components with negligible variance contributions. This process preserves dominant data features while eliminating redundant information, offering two key advantages: (1) substantial reduction in computational resource requirements, and (2) acceleration of downstream analytical tasks via optimized data representations. This synergistic approach effectively balances dimensionality reduction and information retention, addressing both computational efficiency and analytical performance in UML applications.  

In our prior research, \cite{Zhou+2022} developed our initial galaxy morphology classification framework, i.e., the Bagging-based voting clustering approach. By integrating Convolutional Autoencoders (CAE; \citealt{Masci+2011}) to reduce image noise effectively, this method successfully clustered galaxies into 100 groups with high intra-group similarity. Then, \cite{Fang+2023} introduced the Adaptive Polar Coordinate Transformation (APCT) technique, which can improve rotational invariance during image processing, enhancing model robustness and demonstrating GoogLeNet’s superior performance in galaxy classification. Subsequently, \cite{Dai+2023} first systematically combined UML and SML methods to classify 17,292 galaxies in the COSMOS-DASH field. Building on the study of \cite{Dai+2023}, \cite{Song+2024} proposed the \texttt{USmorph} model, which was the same hybrid framework (i.e., UML plus SML) as that in \cite{Dai+2023} but with a self-consistent data pre-processing pipeline, to cluster the galaxy images into 50 groups successfully. This study expanded the dataset sixfold, classified nearly 100,000 galaxies in the COSMOS field, and validated the model’s efficiency and reliability in large-scale classification tasks.

In this work, we introduce an unsupervised clustering algorithm that integrates pre-trained large-scale encoder models with dimensionality reduction techniques. The methodology optimizes computational resource utilization through latent space compression and simplifies feature representations, enabling efficient and streamlined unsupervised clustering.  
Based upon the \texttt{USmorph} framework (\citealt{Song+2024}), we propose two methodological advancements implemented immediately following pre-processing steps:  
(1) we utilize a pretrained ConvNeXt model (\citealt{liu+2022}) to extract hierarchical image features; 
(2) we perform subsequent feature refinement via Uniform Manifold Approximation and Projection (UMAP, \citealt{McInnes+2018}), achieving effective dimensionality reduction.  
The data pre-processing employs a CAE (\citealt{Masci+2011}) combined with APCT (\citealt{Fang+2023}). This step aims to mitigate the effects of noise and enhance rotational invariance in image features.  
Subsequently, the ConvNeXt model is applied to perform deep encoding of preprocessed images, extracting discriminative features. The encoded features undergo dimensionality reduction via the UMAP algorithm, where nonlinear manifold learning further optimizes feature representations. This dual strategy of deep feature extraction and dimensionality reduction is designed to maximize the extraction of critical data features and eliminate redundant dimensions, enabling efficient compression of the feature space.
This not only improves downstream task performance but also significantly reduces computational resource consumption and storage requirements. Finally, a Bagging-based voting clustering model (\citealt{Zhou+2022}) is implemented to classify the refined features. We optimize the UML's predefined cluster number to 20 (reduced from 50 in the original \texttt{USmorph} framework of \cite{Song+2024}, and achieve significant computational savings. The machine recognized 20 pre-stage groups, which were then converged to five physical morphology types. UMAP (\citealt{McInnes+2018}) data visualization and analysis of galaxy morphology parameter distributions are conducted to verify the reliability of the classification results. 

The structure of this paper is as follows. Section \ref{sec:2} provides an overview of the COSMOS program and sample selection criteria. Section \ref{sec:3} introduces the improved unsupervised galaxy morphology classification model, as well as the application of UMAP technology. Section \ref{sec:4} presents classification results and system parameter measurements. Section \ref{sec:5} summarizes the main conclusions and provides an outlook for future research. Throughout this paper, we use the AB magnitude system \citep{Oke+1983} and assume a \cite{Chabrier+2003} initial mass function and a standard flat $\Lambda$CDM cosmology with parameters $H_0 = 70$ km s$^{-1}$ Mpc$^{-1}$, $\Omega_m = 0.3$, and $\Omega_\Lambda = 0.7$.

\section{Observation and Sample Seletction} \label{sec:2}

The COSMOS survey \citep{Scoville+2007} is a large observational program dedicated to studying galaxy evolution, star formation, active galactic nuclei, and the relationship between dark matter and large-scale structure within the redshift range $0.5 < z < 6$. It covers an area of approximately $\rm 2\ deg^2$ and employs multi-wavelength imaging and spectroscopic observations spanning from X-ray to radio. This survey includes high-resolution imaging data from the Hubble Space Telescope (HST) using the Advanced Camera for Surveys (ACS) in the F814W filter, which covers the largest contiguous area ($\sim$1.64 $\rm deg^2$) within the COSMOS region. 
The original image collection comprises approximately 590 pointings, with an average exposure time of 2028 seconds per pointing.
The original images were processed by \citet{Koekemoer+2007} using the \texttt{MultiDrizzle} package \citep{Koekemoer+2003}. The final mosaic images have a pixel scale of $0\farcs03$ and a $5\sigma$ depth of 27.2 AB magnitude for point-source observations within an aperture diameter of $0\farcs24$. Subsequent analyses in this study are based on these high-resolution mosaic images.

The dataset used in this study is from the ``Farmer'' COSMOS2020 catalog \citep{Weaver+2022}, which provides comprehensive photometric data across 35 bands spanning from ultraviolet to mid-infrared wavelengths. The catalog was generated by modeling and performing photometric measurements on multi-band images using the {\tt Tractor package} \citep{lang+2016}. 
Thus, many physical properties can be derived by fitting the spectral energy distribution (SED of galaxies, including redshifts, stellar masses, and other parameters.
For the determination of photometric redshifts, \citet{Weaver+2022} utilized two distinct codes: {\tt EAZY} \citep{Bramme+2008} and {\tt LePhare} \citep{Ilbert+2009}. The redshift estimation process in {\tt LePhare} involved a library composed of 33 galactic templates. A detailed comparison demonstrated high consistency between the two methods in calculating photometric redshifts. In this work, we adopted the redshifts obtained from {\tt LePhare}, as shown in Figure~15 of \citet{Weaver+2022}.

\begin{figure}
    \includegraphics[width=0.45\textwidth]{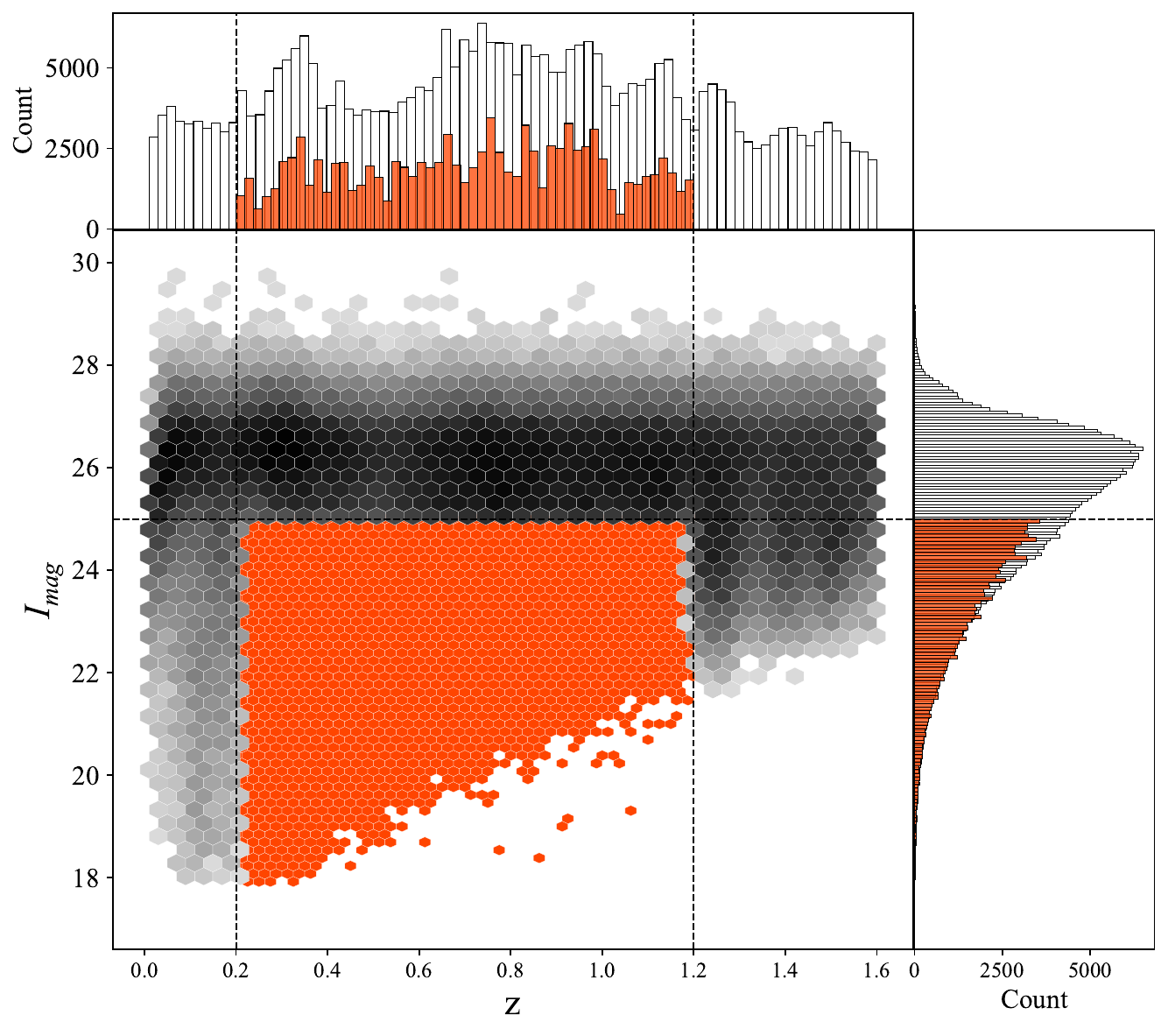}
   \caption{Distribution of galaxies in the COSMOS field in the I-band magnitude-redshift plane. The corresponding number distributions along $I_{\rm mag}$ and redshift are displayed at the top and right corners, respectively. The sample of galaxies at $0.2<z<1.2$ with $I_{\rm mag}<25$ is shown in orange. }   
\label{fig:1}
\end{figure}

Based on the COSMOS2020 catalog, we selected the parent sample of galaxies by below criteria:
(1) $\rm lp_{type}=0$, to ensure that we select galaxies rather than stars;
(2) $\rm FLAG_{ COMBINE}=0$, to make sure that flux measurements are unaffected by the bright stars and that the objects are not at the image edges, which can guarantee reliable measurements on redshifts and stellar mass;
(3) $0.2 < z < 1.2$, to ensure the galaxy morphologies measured in the rest-frame optical;
(4) the high quality of source images with the signal-to-noise ratio (S/N) larger than 5 and without anomalous pixels.
As a result, there are 99,806 galaxies retained. Figure~\ref{fig:1} illustrates the distribution of galaxies between I-band magnitude and redshift in the COSMOS field, where the parent sample is shown in orange, satisfying $0.2<z<1.2$ and $I_{\rm mag}<$25.

\section{METHOD FOR MORPHOLOGICAL CLASSIFICATION} \label{sec:3}
In this section, we present the framework of the updated UML technology, which is structured into four primary components, including data pre-processing, feature extraction, UMAP dimensionality reduction, and voting based on a Bagging clustering model.

\begin{figure*}
    \includegraphics[width=2\columnwidth]{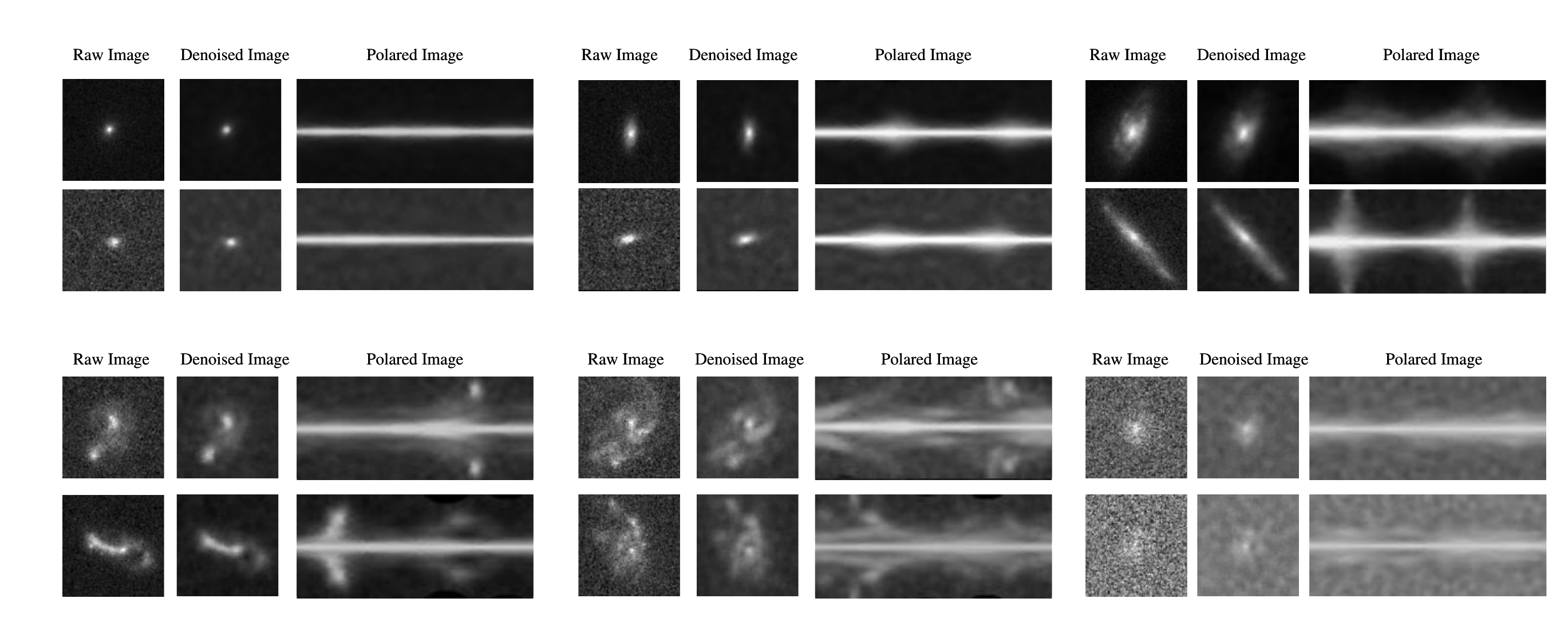}
    \caption{Six examples of the image pre-processing flow, corresponding to each set of images from left to right: original image, denoised image, and image after the polar coordinate transformation.}
    \label{fig:2}
\end{figure*}

\begin{figure}
\centering
  \includegraphics[scale=0.40]{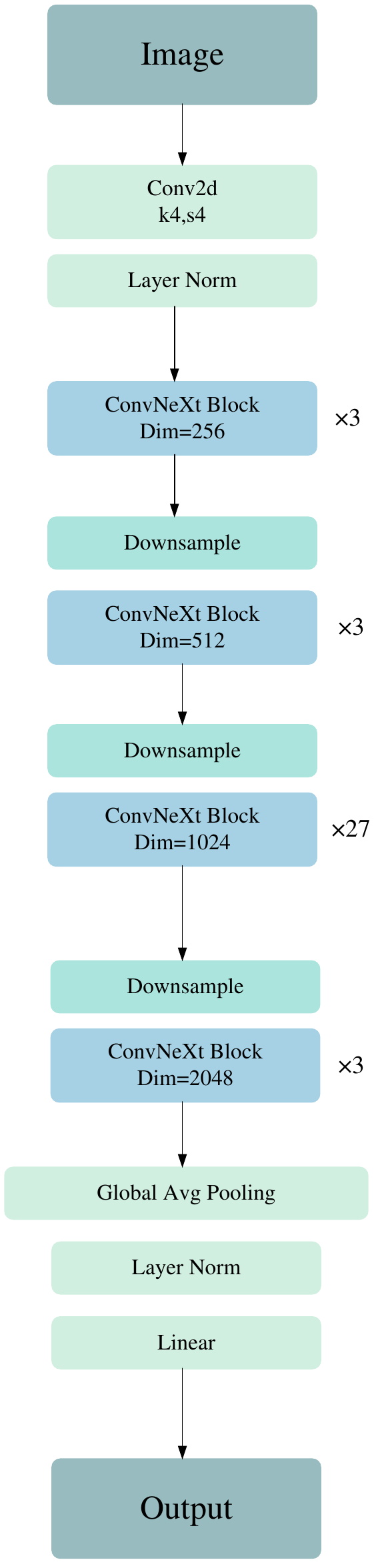} 
  \caption{Framework of ConvNeXt model. The architecture of ConvNeXt employs a Transformer-inspired modular design, which hierarchically stacks standardized modules in a multi-level architecture. Inter-module communication and gradient propagation are facilitated via consistent architectural interfaces, ensuring efficient information flow and stable training dynamics. This design paradigm maintains structural simplicity while enhancing scalability, enabling systematic network expansion via modular composition.}
\label{fig:3}
\end{figure}  

\subsection{Data pre-processing}
During the data pre-processing stage, we uniformly cropped all images to $100 \times 100$ pixels to ensure that sufficient structural information is preserved. As we find that the cropped size is related to the effective radii of galaxies in our sample, we adopt the \texttt{GALAPAGOS} software (\citealt{Barden+2012,haubler+2022}) to measure the effective radius of each galaxy. The following statistical analysis reveals that 98\% of the sample galaxies have effective radii smaller than 50 pixels, suggesting that the adopted pixel size (i.e., $100 \times 100$ pixels) used to crop images is enough to support subsequent morphological classification tasks.

To enhance the model’s classification stability, we incorporate data pre-processing steps. In response to the adverse effects of high noise levels -- resulting from low S/N -- on galaxy classification accuracy, we apply CAE for image denoising. During CAE processing, convolutional and pooling operations automatically extracted key features, which were then used to reconstruct denoised images while preserving essential structural information intact. Table \ref{tab1} lists the architecture and layer parameters of the CAE employed in this study. As shown in Figure \ref{fig:2}, the left column of each sample image set represents the original data, and the middle column displays the CAE-denoised reconstructed images. This process effectively removes noise and artifacts from the images while successfully retaining the fundamental morphological features of the galaxies.  

In galaxy morphology classification, spatial rotations can lead to misclassification of morphological types. To eliminate rotational interference and achieve rotation-invariant feature extraction, we implement the APCT technique proposed by \cite{Fang+2023} to enhance the model robustness by constructing a rotation-invariant feature space. The workflow of this method involves three steps. First, an initial polar axis should be defined by connecting extremum brightness points (maximum/minimum pixel values) in the image. Subsequently, the polar axis needs to be rotated counterclockwise in increments of 0.05 radian, where pixels along the rotational polar axis are aggregated into polar coordinates via a stacking process. Finally, we apply mirror transformations to accentuate central features. 

By employing an APCT, image rotations are converted into translations in the polar coordinate system, thereby transforming the inherent translation invariance of CNNs into rotation invariance. 
As demonstrated by \citet{Fang+2023}, APCT maintains classification accuracy at a level comparable to unrotated images even under large rotations (90\textdegree--270\textdegree), whereas models without APCT suffer a substantial performance drop. This clearly highlights its effectiveness in improving orientation invariance.
Compared with conventional rotation-based data augmentation (e.g., \citealt{Reyes+2018, Yao+2019}), APCT achieves rotation invariance more efficiently, thereby reducing the computational burden in large-scale sky survey analyses.
Furthermore, APCT processing accentuates feature contrasts and improves label alignment, leading to further gains in classification accuracy. The right panel of each image set in Figure \ref{fig:2} displays the corresponding results after APCT implementation.  

\begin{table}
\center
\caption{The CAE architecture illustration} \label{tab1}
\begin{tabular}{ccccc}
\hline\hline
Layer & Operation & Dimension & Filter Size  & Stride \\
\hline
L0 & Input       & $100\times 100 \times 1$ & ... & ... \\
L1 & Convolution & $100\times 100 \times 8$ & $5\times 5$ & ... \\
L2 & Maxpooling  & $50 \times 50  \times 8$ & $2\times 2$ & $2\times 2$ \\
L3 & Convolution & $50 \times 50  \times 8$ & $5\times 5$ & ... \\
L4 & Maxpooling  & $25 \times 25  \times 8$ & $2\times 2$ & $2\times 2$ \\
L5 & Unfolding   & 10000   & ... & ... \\
L6 & Full connection & 40 & ... & ... \\
\hline
\end{tabular}
\begin{tablenotes}
\item[]{\footnotesize{Note: The CAE employs a symmetric encoder-decoder architecture. The encoder processes a $100\times100$ pixel input through two convolutional layers. The first layer has 16 channels with $5\times5$ kernels and max-pooling ($2\times2$ kernel, stride 2). The second layer uses 8 filters and max-pooling ($2\times2$ kernel, stride 2), resulting in $25\times25\times16$ feature maps. These are flattened and passed through a fully connected layer to produce a 40-dimensional latent space. The decoder reconstructs the output using unpooling and deconvolution to mirror the encoder structure.}}
\end{tablenotes}

\end{table}

\begin{figure*}
 \centering 
\includegraphics[width=0.70\textwidth]{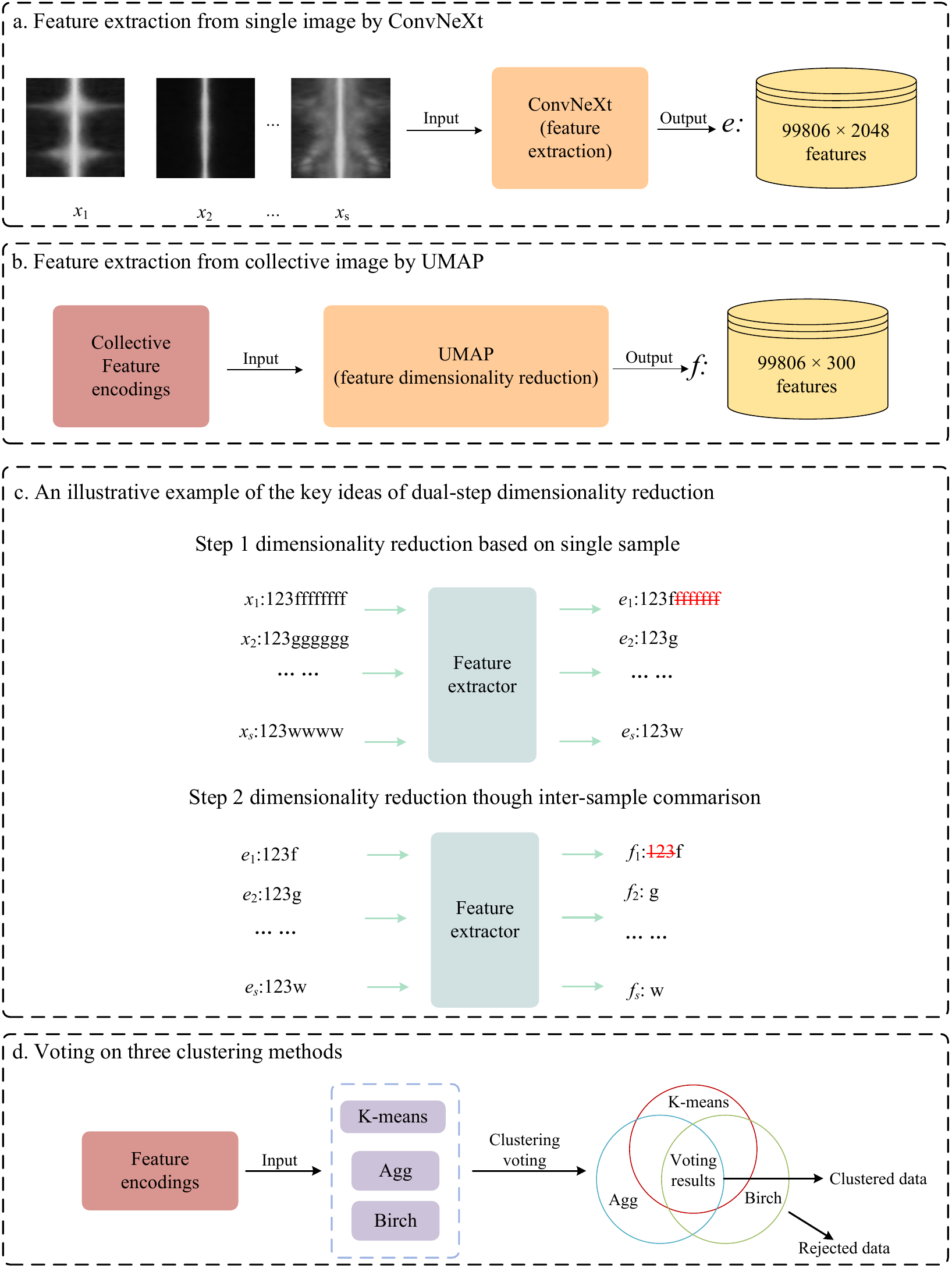} %
    \caption{Schematic diagram of UML clustering process, including to extract key features from image data using ConvNeXt in step (a); to reduce dimensionality and remove redundant image information via UMAP in step (b); to provide a detailed example of the feature extraction and dimensionality reduction process in step (c); and to adopt a voting based on Bagging clustering model mechanism in step (d).
}
    \label{fig:4}
    
\end{figure*}

\subsection{ConvNeXt model framework and feature extraction}
In recent years, deep visual models have shown remarkable capability in learning hierarchical representations from large-scale datasets (e.g., \citealt{Krizhevsky+2012}). Pre-trained architectures can serve as powerful feature extractors, enabling the extraction of transferable representations directly from unlabeled data.
These models can effectively extract discriminative features without requiring labels for unlabeled data. ConvNeXt (\citealt{liu+2022}) is an efficient visual architecture. It enhances the ResNet framework (\citealt{he2+2015}) by introducing a patchify layer and adopting depthwise grouped convolutions to widen the network. This design merges the strengths of CNNs and Swin Transformer (\citealt{dosovitskiy+2021}), leveraging CNNs' robust local feature extraction while modeling long-range dependencies through self-attention mechanisms. Consequently, ConvNeXt demonstrates superior performance in image classification tasks (e.g.,\citealt{Woo+2023,liu+2023}).
We utilize the ConvNeXt-22k model pre-trained on the ImageNet-22k dataset (\citealt{Russakovsky+2015}). The architecture begins with an initial convolutional layer using a 4$\times$4 kernel and a stride of 4, followed by LayerNorm for normalization. The model comprises multiple stages of ConvNeXt Blocks, which integrate depthwise separable convolutions, LayerNorm, GELU activation, and residual connections to facilitate effective feature extraction and information retention. These blocks are organized into groups with progressively increasing channel dimensions (256, 512, 1024, 2048), interspersed with downsampling layers to adjust spatial resolution. The network concludes with global average pooling, LayerNorm, and a linear classification layer (see Figure \ref{fig:3}).

The ConvNeXt model employs depthwise separable convolutions to construct a lightweight architecture while enabling multi-scale feature extraction.
The polarized image obtained after data processing is input into the ConvNeXt model for feature extraction, and galaxy images are transformed into 2048-dimensional feature vectors $e$, which efficiently filter out redundant information present in the original imagery, as illustrated in the corresponding step of Figure \ref{fig:4}.

\subsection{UMAP dimensionality reduction}\label{sec:3.3}
To further extract key features, we employ dimensionality reduction techniques to compress the 2048-dimensional feature vectors output by the ConvNeXt model, optimizing the feature space while preserving essential information. Among traditional linear dimensionality reduction methods, PCA achieves dimensionality reduction by identifying orthogonal directions of maximal variance in high-dimensional data (principal components), maintaining the global linear structure.
However, its linear nature imposes significant limitations when handling nonlinear manifold structures: when data distributions exhibit complex nonlinear patterns, PCA fails to effectively capture nonlinear relationships among data points and retains the local structure of nonlinear data poorly, necessitating the retention of excessive principal components to maintain information integrity, thereby drastically reducing reduction efficiency (e.g., \citealt{McInnes+2018,gong+2018,liu+2023}).

For high-dimensional data such as images, which often exhibit nonlinear manifold structures, the feature vectors extracted by pretrained encoders typically reside on a low-dimensional manifold (e.g., \citealt{Mao+2024,lowe+2024}). As a nonlinear method rooted in manifold learning and topological principles, UMAP approximates the underlying manifold structure of data and preserves its topological relationships, enabling more precise revelation of intrinsic geometric features. Compared with linear dimensionality reduction algorithms, UMAP achieves significant dimensionality reduction while effectively balancing global and local structures and maintaining high information integrity, particularly excelling in handling high-dimensional data with complex nonlinear distributions, such as images. Specifically, UMAP's dimensionality reduction process comprises three key steps: a) quantifying local similarities between each high-dimensional data point and its neighborhood points using an exponential probability distribution; b) strategically constructing a minimum spanning tree from these high-similarity connections to capture the global topological structure of the data manifold; c) employing stochastic gradient descent optimization to iteratively refine the low-dimensional embedding space by minimizing the cross-entropy loss between high-dimensional and low-dimensional topological representations. Despite the complexity of its topological construction and optimization procedures, UMAP systematically reduces the computational complexity of key steps to near-linear levels. This is achieved through the integration of approximate nearest neighbor search, sparse graph representations, and optimization strategies that avoid global normalization (\citealt{McInnes+2018}). Consequently, UMAP achieves significantly higher efficiency compared to other nonlinear dimensionality reduction algorithms, making it well-suited for processing future large-scale sky survey data. This process systematically reduces discrepancies between high-dimensional and low-dimensional topological characterizations, thereby achieving efficient dimensionality reduction while preserving the essential features of the data.

In nonlinear dimensionality reduction tasks, the selection of the low-dimensional space dimensionality in UMAP impacts algorithmic performance. Unlike linear methods such as PCA, which can determine optimal dimensionality through variance explained ratio calculations using the “elbow method”, UMAP primarily aims to preserve both local and global geometric structures of data rather than adhering to variance maximization principles. This necessitates manual specification of target dimensionality during the reduction process. Current research predominantly relies on empirical approaches or downstream task performance evaluation to select the dimension, which is highly subjective and computationally expensive
(e.g., \citealt{liu+2023,ge+2023}). Consequently, the development of appropriate dimensionality selection strategies becomes crucial for enhancing dimensionality reduction efficacy.

The Davies-Bouldin Index (DB Index; \citealt{Davies+1979}) is an internal metric for evaluating clustering quality. As an unsupervised measure, it operates independently of external information and assesses the validity of clustering results by quantifying the ratio between intra-cluster compactness and inter-cluster separation. The DB Index can be expressed as:
\begin{equation}    
\mathit{DB}=\frac{1}{N}\sum_{i=1}^N \max_{j \neq i} \left( \frac{S_i+S_j}{d(c_i,c_j)} \right);
\end{equation}
where $S_i$ and $S_j$ are the cohesiveness of cluster $i$ and cluster $j$, respectively, defined as the average distance from the data points within the cluster to the cluster center. $d(c_i, c_j)$ represents the distance between the centers of cluster $i$ and cluster $j$. $N$ is the number of clusters, and $c_i$ and $c_j$ are the centers of the clusters.
The DB index has a significant advantage in handling convex-shaped clusters \citep{Halkidi+2001}.
The smaller the index value, the better the clustering performance. To achieve a balance between effectiveness and efficiency in the UMAP dimensionality reduction process, we compute the DB index across dimensionality values ranging from 50 to 500 in increments of 50. The results show that the DB index reaches its minimum value at 300 dimensions. Consequently, we select 300 dimensions as the optimal output for dimensionality reduction. 
When determining the target dimensionality, the DB index is computed based on the clustering result from K-means. The calculation inherently leverages the cluster centroids to quantify intra-cluster dispersion and inter-cluster separation. This evaluation was performed with the number of clusters fixed at $k=$20. For a comprehensive explanation of the clustering methodology and the rationale behind the choice of cluster count, please refer to Section \ref{sec:3.4}.

As illustrated in Figure \ref{fig:4}, after applying UMAP for dimensionality reduction, each galaxy image \( x \) is compressed from its original 2048-dimensional representation \( e \) (see panel a) to a 300-dimensional representation \( f \) (see panel b). In the two-step feature extraction process (see panel c), the first step removes redundant features inherent in the data, retaining only the valid features. The second step further eliminates redundant features by comparing the differences between samples, and highlights key features.

\begin{figure*}
    \centering
    \includegraphics[width=0.9\textwidth]{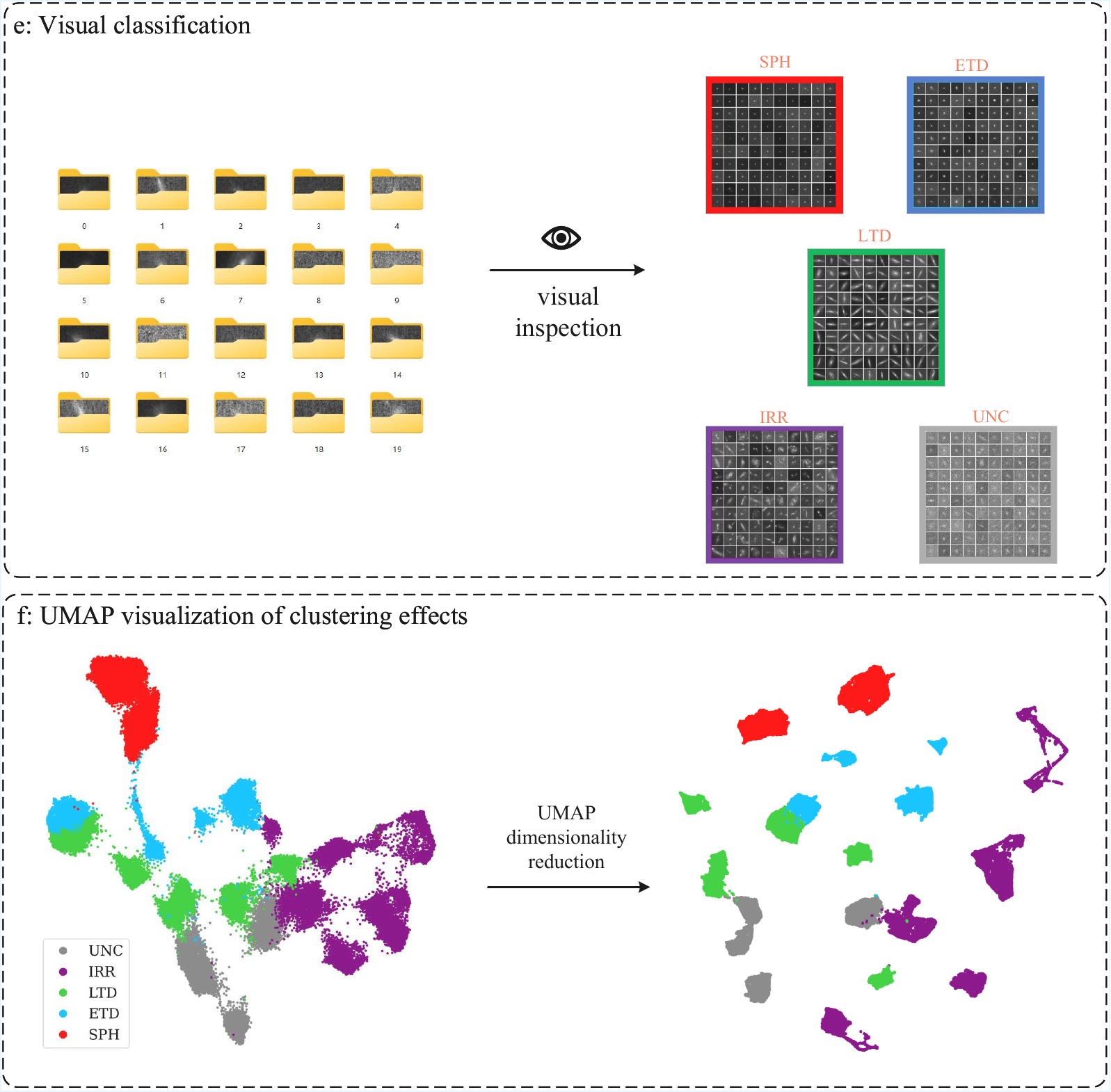}
    \caption{The continued schematic diagram of the UML clustering process as in Figure~\ref{fig:4}. Step (e) displays the Visual classification by randomly selecting 100 images from the 20 machine-learning clusters and visually classifying them into five types of galaxies, including SPH (spherical), ETD (early-type disk), LTD (late-type disk), IRR (irregular), and UNC (unclassified), respectively. Step (f) shows the UMAP visualization of clustering effects by analyzing the UMAP 2D projection of the five-class labels derived from the previous step. On the left, visualize the 2048-dimensional features of all samples extracted by the ConvNeXt model using UMAP; on the right, visualize the distribution of 300-dimensional features after UMAP-based dimensionality reduction.}
    \label{fig:5}
\end{figure*}

\begin{figure*}
\includegraphics[width=2\columnwidth]{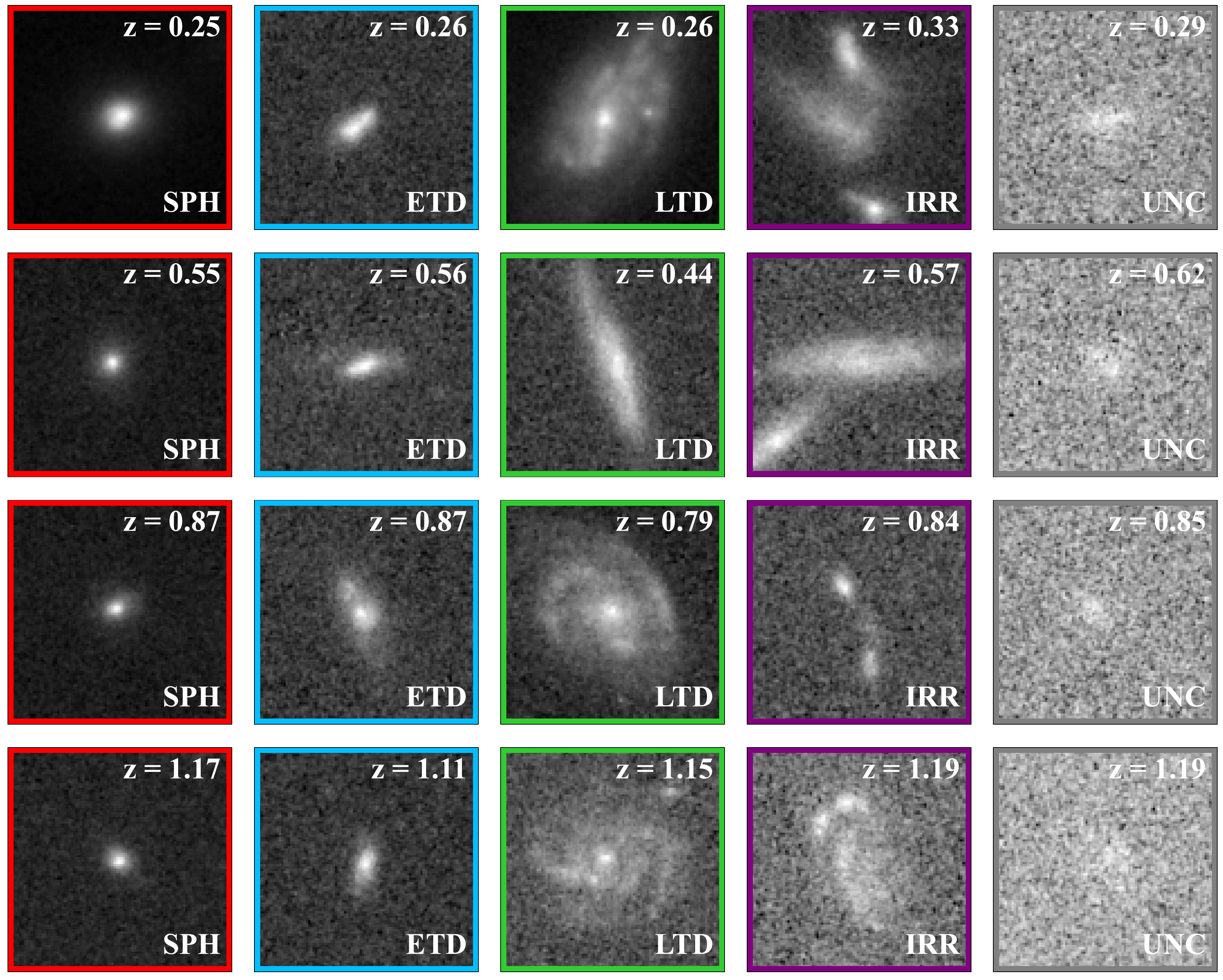}
    \caption{Example images of galaxies randomly selected from the sample. From left to right, images in each column correspond to spherical galaxies (SPHs), early-type disk galaxies (ETDs), late-type disk galaxies (LTDs), irregular galaxies (IRRs), and unclassified galaxies (UNCs), respectively.}
    \label{fig:6}
\end{figure*}

\begin{figure*}
\centering
\includegraphics[width=1.5\columnwidth]{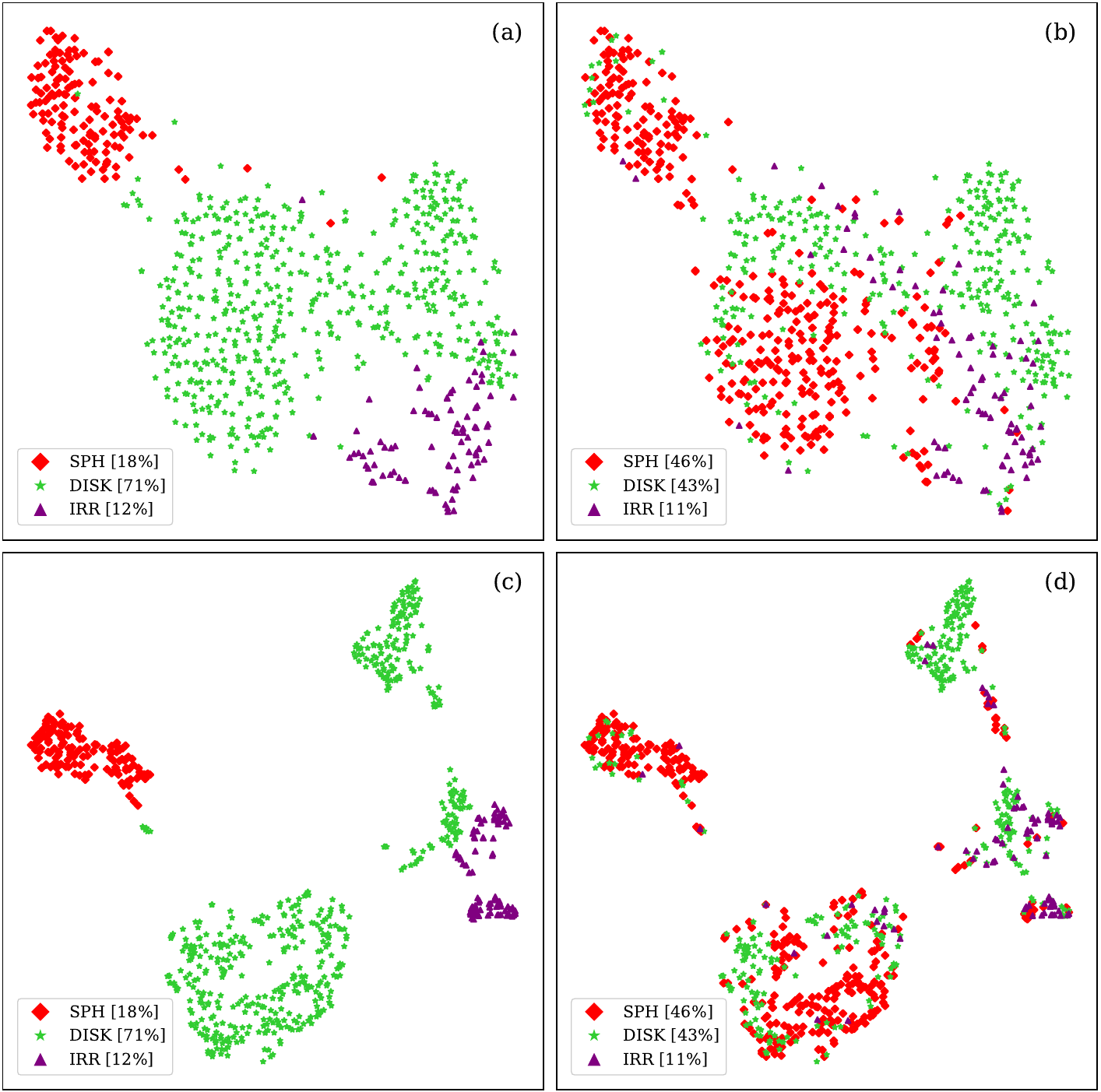}
    \caption{Two-dimensional visualization of morphological classification results for the matched subsample. Panel (a) depicts the t-SNE projection of the proposed UML method on ConvNeXt features (2048-dimensional), while Panel (b) presents the corresponding results from the \texttt{Galaxy Zoo: Hubble} catalog (\citealt{Willett+2017}) on the same features. Panels (c) and (d) display t-SNE projections of the two methods on 300-dimensional features, respectively. Comparative analysis demonstrates that the proposed UML method (Panels a and c) exhibits more compact intra-class clustering and more distinct inter-class separation.}
    \label{fig:7}
\end{figure*}

\subsection{UML Clustering Process}\label{sec:3.4}
After data pre-processing, feature extraction, and UMAP dimensionality reduction, We employ a bagging-based multi-model voting classification method (\citealt{Zhou+2022}) to improve classification reliability. This approach can remove sources with inconsistent cluster memberships across different algorithms. In this process, we use three algorithms to perform clustering voting: the Hierarchical Balanced Iterative Reduction and Clustering Algorithm (Birch; \citealt{Zhang+1996}), the K-means Clustering Algorithm (K-means; \citealt{Hartigan+1979}), and the Agglomerative Clustering algorithm (Agg; \citealt{Murtagh+1983, Murtagh+2014}). Each algorithm processes the data independently by partitioning the sample into 20 groups. We use the labels generated by the K-means algorithm as the reference labels, assign the corresponding labels to the other clustering models by counting the frequency of distribution of K-means labels within each group, and synchronize the majority voting mechanism. When the voting results of the three algorithms are identical, the samples that they contain together will be determined as the final sample set, and the rest of the samples that are not included in this set will be excluded. The clustering process refers to Step d in Figure~\ref{fig:4}.

In unsupervised clustering, the number of clusters \(k\) is typically set to a relatively high value based on the scientific objectives. A small \(k\) may fail to capture subtle differences in the data, resulting in overly coarse groupings, whereas an excessively large \(k\) can lead to overfitting, amplify noise, increase computational cost, and produce unstable results. Specifically, a low \(k\) (e.g., $k=$ 5) fails to resolve key morphological classes, potentially causing galaxies to be grouped by non-morphological features such as S/N, while a high \(k\) (e.g., $k=$ 50) produces redundant clusters, increasing computational load and drastically reducing the efficiency of subsequent visual inspection and labeling. To balance performance and efficiency, we adopt \(k = 20\) as our clustering number. This choice effectively groups galaxy images with similar features while avoiding an excessive burden for subsequent manual labeling, thereby achieving a trade-off between classification accuracy and operational efficiency.

Based on the 300-dimensional feature representation obtained through UMAP dimensionality reduction, clustering was performed using a Bagging-based approach. After excluding 49,750 samples (49\%) with inconsistent votes, a total of 50,056 galaxies were successfully classified into 20 distinct groups. To further analyze group classifications and assess differences in physical properties, a visual classification process was conducted. For this purpose, 100 images were randomly selected from each group, and three trained experts participated in the visual inspection step to minimize subjective bias. Final classifications were assigned only when at least two experts reached agreement. 
Consistent with our over-clustering strategy, the majority of galaxies within each group display highly similar morphological features. Empirically, more than 80\% of the galaxies in each cluster were found to be morphologically homogeneous, which ensured that the 100 randomly inspected images provided a representative characterization of the entire cluster.
This workflow is illustrated in Step e of Figure~\ref{fig:5}. Ultimately, these 50,056 galaxies were categorized into five morphological classes: spherical (SPH), early-type disk (ETD), late-type disk (LTD), irregular (IRR), and unclassified (UNC). The distribution of the number of galaxies in each category is detailed in Table \ref{tab2}.

\begin{table}
\center
\caption{Number of galaxies in different morphological types\label{tab2}}
\begin{tabular}{ccccccc}
\hline\hline
Model & SPH & ETD & LTD  & IRR & UNC& TOTAL\\
\hline
UML & 8,506 & 7,348 & 11,473 & 14,044 & 8,685 & 50,056\\
\hline
\end{tabular}
\end{table}

\section{RESULTS AND DISCUSSION}\label{sec:4}
\subsection{Overall morphological classification results}

Notably, UMAP \citep{McInnes+2018} is not only widely applied in dimensionality reduction and clustering tasks but also frequently used for visual inspection of high-dimensional data. This algorithm efficiently maps high-dimensional complex data to a low-dimensional space (e.g., 2D or 3D) while preserving both local similarities and global structural characteristics. Thus, we utilize UMAP to project the 50,056 galaxy datasets into a 2D plane for intuitive visualization and analysis.

As shown in step f of Figure~\ref{fig:5}, the 2048-dimensional feature vectors extracted by the ConvNeXt model exhibit non-convex distribution characteristics due to their high dimensionality. After UMAP dimensionality reduction, the data becomes to exhibit convex clusters, with 90\% of galaxy samples showing no overlap in the projection and clear boundaries between galaxy groups. 
Through updated UML technology, we have successfully classified 50,056 galaxies (51\%) in the COSMOS field. This includes 8,506 SPHs, 7,348 ETDs, 11,473 LTDs, 14,044 IRRs, and 8,685 UNC. Figure \ref{fig:6} shows sample images of some galaxies. Intuitive comparisons reveal distinct differences between them.

\subsection{Validation through Comparison with External Morphology Catalogs}
To assess the scientific reliability of our enhanced UML-based automated classification method, we compare its results with those from \texttt{Galaxy Zoo:Hubble }(GZH; \citealt{Willett+2017}). The GZH project, which relies on visual classifications by volunteers, adopts a classification framework centered on three principal structural types: SPH, DISK, and IRR. It produces a set of probabilistic morphological features, such as smoothness and the presence of structures or clumps, and develops a scalable automated classifier through statistical learning techniques. 
Since the GZH project offer
adjustable thresholds to balance sample completeness and classification confidence, combined with HST/ACS F814W data from the COSMOS field, it can serve as a suitable  reference dataset for this work.

To obtain robust and reliable galaxy morphologies, we applied the following stringent thresholding criteria (see \citealt{Willett+2017}) to map the probability components in GZH to three categories (SPH, DISK, and IRR). The specific thresholds used are as follows:
{\raggedright
\begin{enumerate}
  \item SPH: \( f_{\text{smooth, best}} > 0.70 \), \( f_{\text{features, best}} < 0.30 \), and \( f_{\text{round}} > 0.50 \);
  \item DISK: \( f_{\text{features, best}} > 0.23 \), \( f_{\text{clumpy, no}} > 0.30 \), \( f_{\text{edge-on, no}} > 0.25 \), \( f_{\text{spiral}} > 0.25 \), \( f_{\text{odd, yes}} < 0.30 \), and \( f_{\text{irregular}} < 0.30 \);
  \item IRR: (\( f_{\text{features, best}} > 0.23 \), \( f_{\text{clumpy, yes}} > 0.80 \), \( f_{\text{clump, asym}} > 0.50 \)) or (\( f_{\text{odd, yes}} > 0.50 \), and \( f_{\text{irregular}} > 0.50 \)).
\end{enumerate}
}
By cross-matching our sample with a curated morphological catalog, we ultimately obtained 757 galaxies with stable morphologies for subsequent comparison. During this process, the UNC category was excluded as its proportion fell below 1\% after matching.

Using the GZH catalog as a reference, we first compared the direct classification results of the two schemes. The overall consistency rate was 58\%, indicating that although both methods captured similar global trends, notable discrepancies persisted near the morphological boundaries. To further investigate these differences, we performed a visual analysis in the feature space, adopting the t-SNE algorithm (\citealt{van+2008}) because it is more effective than UMAP at preserving local neighborhood structures in small samples (fewer than 1,000 galaxies). The features were derived from the ConvNeXt model and included both the original 2048-dimensional representations extracted directly from galaxy images and the 300-dimensional representations obtained via UMAP dimensionality reduction. By employing this dual-feature strategy, both sets of galaxy labels were projected into the same feature space derived from the original image representations, ensuring a fairer comparison.

As shown in Figure~\ref{fig:7}, both in the original 2048-dimensional feature space and the reduced 300-dimensional space, our unsupervised classification results exhibit tighter intra-cluster distributions and clearer inter-cluster separation in the projection space. Compared to the distributions obtained through GZH probability label mapping, our approach demonstrates higher consistency and improved separability.

\subsection{Validation with Galaxy Morphology and Structure Parameters}
Since UMAP visualization only provides qualitative insights, we will conduct quantitative validation of the classification results by analyzing the distributions of multiple morphological parameters in this section. 
In general, parametric and nonparametric structural parameters serve as fundamental methods for characterizing their morphological classifications, so that galaxies with different morphologies typically exhibit specific distribution patterns in parameter space (e.g., \citealt{Zhou+2022, Dai+2023, Yao+2023, Song+2024}).

Due to the poorly understood evolutionary mechanisms of low-mass galaxies and to ensure the accuracy and reliability of physical parameter measurements, our analysis focuses exclusively on massive galaxies with stellar masses exceeding $10^{9}~M_{\odot}$. Out of the 50,056 successfully classified galaxies, 21,475 (42.9\%) galaxies have stellar masses below this threshold and are therefore excluded, while 24,380 (48.7\%) massive galaxies (after removing UNC-type) are retained for subsequent analysis in this subsection. We exclude UNC-type galaxies because their morphological parameters are difficult to measure, primarily constrained by relatively weak S/N.

\begin{figure*}
    \includegraphics[width=2\columnwidth]{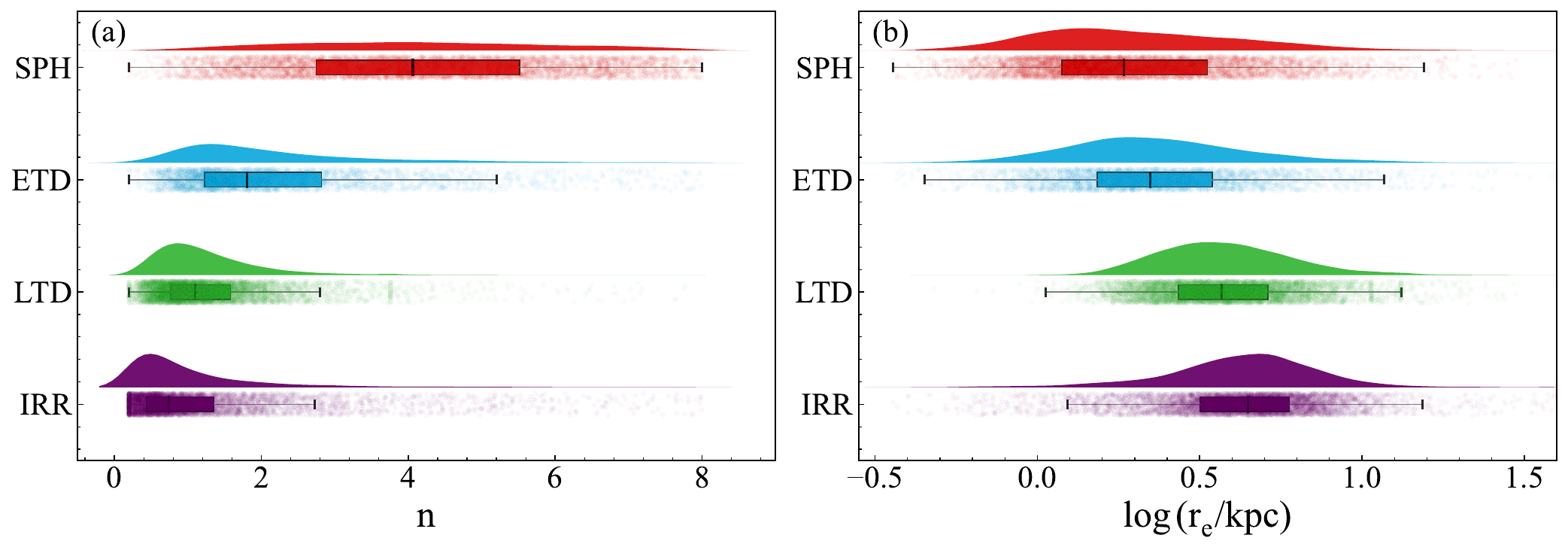}
    \caption{Raincloud plots of the S\'{e}rsic index (left panel) versus effective radius (right panel) for different massive galaxy types. The ``cloud'' portion of the plot reflects the data density distribution characteristics, and the ``rain'' plot portion demonstrates the degree of dispersion of the sample values. In the box plot, the lower boundary corresponds to the first quartile, the upper boundary corresponds to the third quartile, the horizontal line in the center indicates the median, and the upper and lower whisker ends of the box represent the maximum and minimum values. It can be seen that the S\'{e}rsic index gradually decreases as the galaxy type goes from SPHs to IRRs, while the effective radius gradually increases.}
    \label{fig:8}
\end{figure*}

\begin{figure*}
    \includegraphics[width=2\columnwidth]{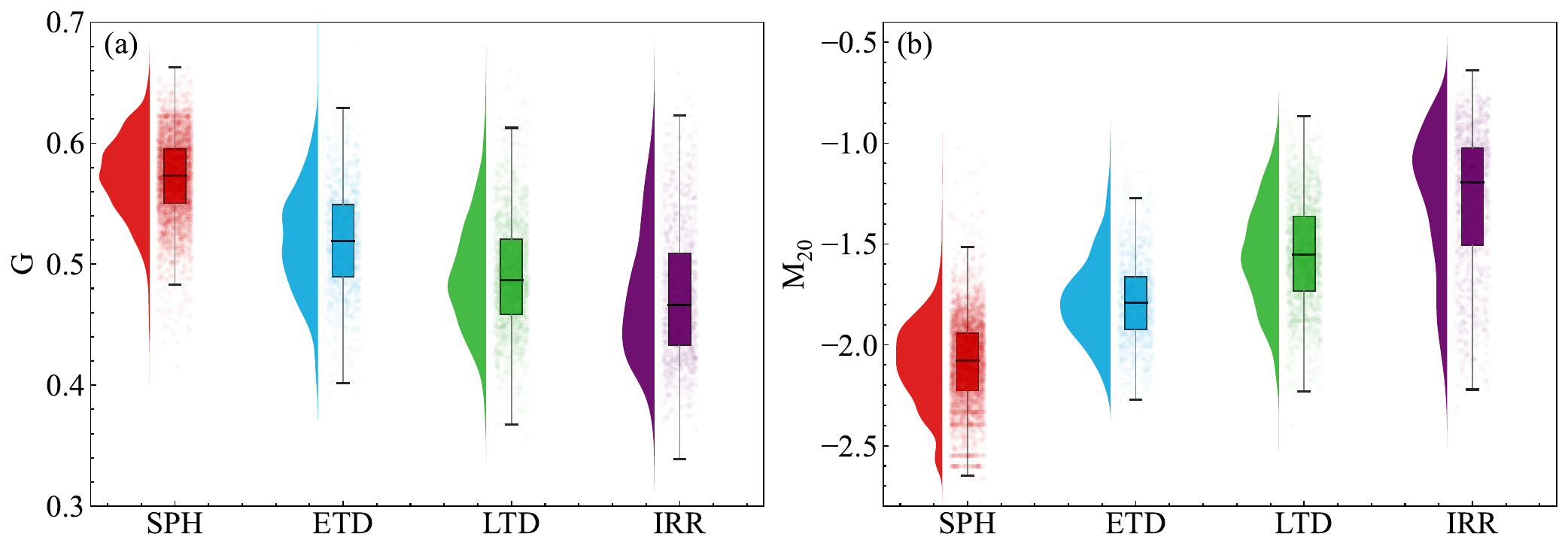}
    \caption{Raincloud plots of $G$ (left panel) and $M_{20}$ (right panel) for different types of massive galaxies. Analogues are similar to Figure~\ref{fig:8}. It can be seen that from SPHs to IRRs, the $G$ of galaxies gradually decreases, while the $M_{20}$ gradually increases.}
    \label{fig:9}
\end{figure*}

\begin{figure*}
    \includegraphics[width=2\columnwidth]{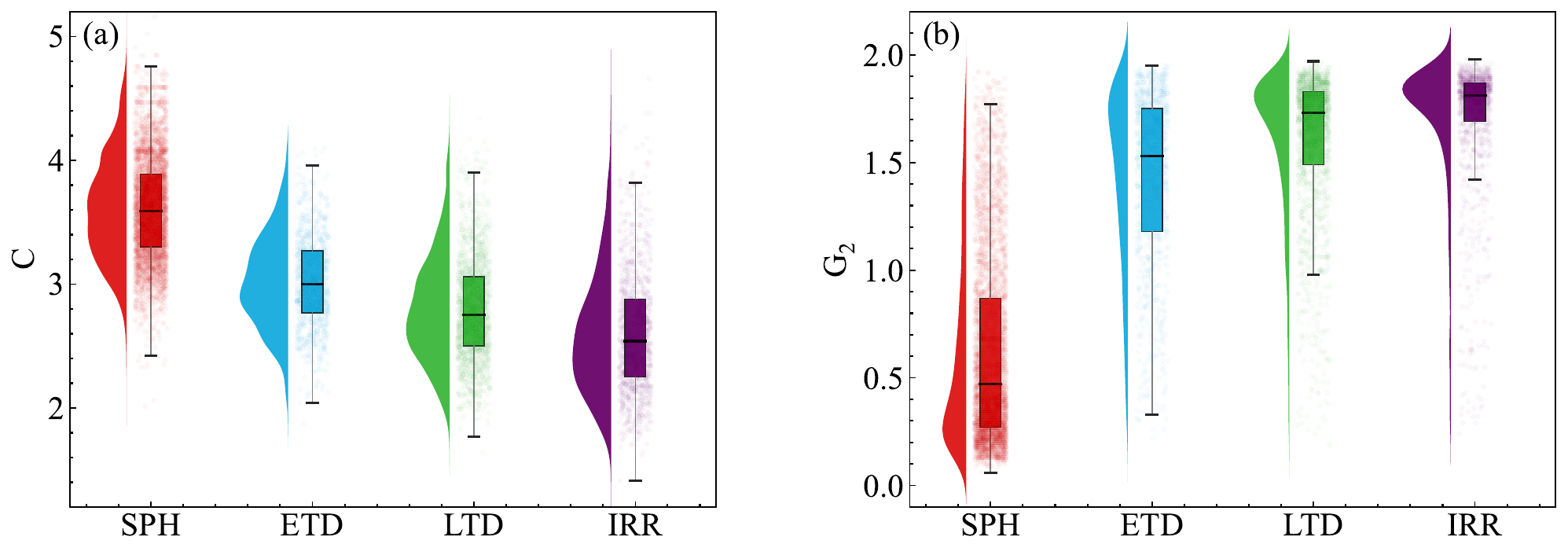}
    \caption{Raincloud plots of $C$ (left panel) and $G_{2}$ (right panel) for different types of massive galaxies. Analogues are similar to Figure~\ref{fig:8}. It can be seen that from SPHs to IRRs, the $C$ of galaxies gradually decreases, while the effective $G_{2}$ gradually increases.}
    \label{fig:10}
\end{figure*}

\begin{figure*}
    \includegraphics[width=2\columnwidth]{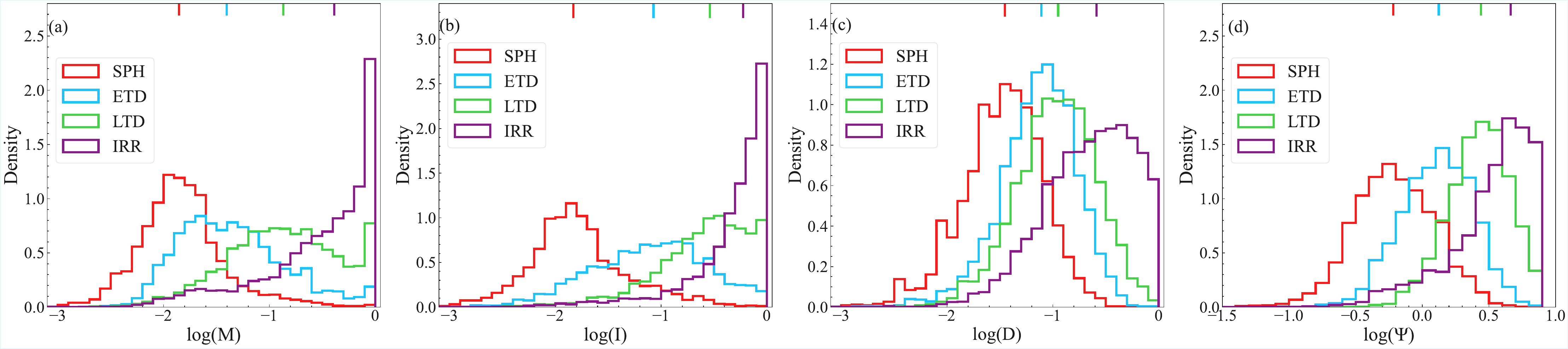}
    \caption{Distributions of Multimode ($M$, panel a), Intensity ($I$, panel b), and Deviation ($D$, panel c), and multiplicity (\textit{$\Psi$}, panel d) for different types of massive galaxies. The top bars represent the median values of $M$, $I$, $D$,  and \textit{$\Psi$} for different types of galaxies. The median values of the galaxies are gradually increasing from SPHs to IRRs.}
    \label{fig：11} 
\end{figure*}

\begin{figure*}
    \includegraphics[width=2\columnwidth]{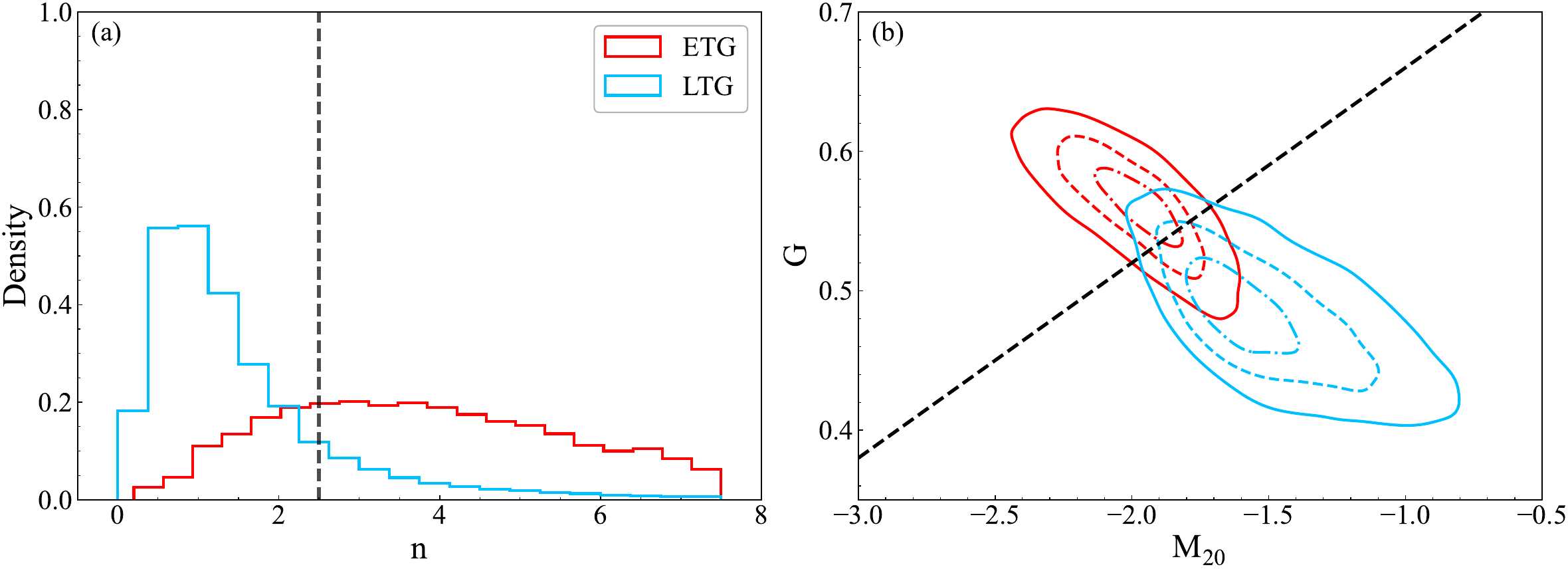}
    \caption{The panel (a) displays the distributions of the S{\'e}rsic index for early-type galaxies (ETGs) and late-type galaxies (LTGs). The gray dashed line marks the classification threshold at \( n = 2.5 \).  The panel (b) shows the distribution in the \( G - M_{20} \) plane for galaxies classified as ETGs and LTGs. The contour levels enclose 20\%, 50\%, and 80\% of the respective subclass, from the innermost to the outermost regions. The dashed line, given by \( G = 0.14 M_{20} + 0.80 \) (\citealt{Lotz+2008}), represents the boundary between LTGs and ETGs.}
    \label{fig：12} 
\end{figure*}

\subsubsection{Parametric Measurements} 
To obtain the morphological parameters of galaxies, we use the {\tt GALFIT} package \citep{Peng+2002} and the {\tt GALAPAGOS} software \citep{Barden+2012, haubler+2022} to fit the surface brightness profiles of galaxies with a single S\'{e}rsic model. 
We derive the S\'{e}rsic index $n$ and effective radius $r_e$%
\footnote{Defined as the radius enclosing half of the galaxy's total luminosity.} for each galaxy.

In Figure \ref{fig:8}, panel (a) shows the distribution of S\'{e}rsic indices for the four galaxy types, with the median values for SPHs, ETDs, LTDs, and IRRs being 4.06, 1.81, 1.10, and 0.74, respectively. In general, SPHs tend to have higher $n$ values due to their more concentrated light profiles, while IRRs typically have lower $n$ values because of their more diffuse light distributions. As the light concentration decreases, the $n$ values show a decreasing trend from SPHs to IRRs.
Panel (b) illustrates the distribution of effective radii for the four galaxy types, with median $r_e$ values of 1.86, 2.24, 3.71, and 4.47 kpc for SPHs, ETDs, LTDs, and IRRs, respectively. Typically, SPHs are more compact in structure and thus have smaller $r_e$, while  IRRs are more diffuse and exhibit larger $r_e$ values. As the spatial extension increases, $r_e$ shows an increasing trend from SPHs to IRRs. In short, the trends in S\'{e}rsic index and effective radius across different galaxy types are consistent with our expectations regarding their correlations.

\subsubsection{Nonparametric Measurements}
High-redshift galaxies exhibit irregular morphological structures, posing significant challenges in fitting their brightness profiles with empirical functions. In contrast, non-parametric morphological measurements require no presumptive light distribution models and can effectively capture structural characteristics of galaxies, making them widely applicable in evolutionary studies of high-redshift galaxies \citep{Ferreira+2023, Treu+2023}.

The most widely used nonparametric morphological indicators are the \textit{CAS} statistics \citep{Conselice+2000, Conselice+2003}, which include three parameters: concentration ($C$), asymmetry ($A$), and smoothness ($S$). Another widely adopted set of nonparametric morphological indicators is the $G$-$M_{20}$ (Gini-moments20; \citealt{Lotz+2004, Lotz+2006}) classification system. These parameters can be measured without requiring predefined galaxy center positions, making them particularly advantageous for studying high-redshift irregular galaxies and merging systems. Recently, \cite{Freeman+2013} introduced the \textit{MID} statistic, comprising three components: multimode ($M$), intensity ($I$), and deviation ($D$). We employ the \texttt{statmorph\_csst} \citep{Yao+2023} package to measure these parameters.
Additionally, we have incorporated extended optional morphological parameters as supplementary indicators, including \textit{$\Psi$} \citep{Law+2007} and $G_{2}$ \citep{Rosa+2018}. Given the widespread application of these parameters in numerous studies, detailed definitions and computational methodologies can be found in the corresponding references.

From the raincloud plot in Figure \ref{fig:9}, we analyze the trends of the $G$ and $M_{20}$ parameters for different types of galaxies. Specifically, the $G$ coefficients of galaxies show a gradual decrease from SPHs to IRRs, with median values of 0.57, 0.52, 0.49, and 0.47, respectively. Conversely, the $M_{20}$ values of galaxies exhibit a gradual increase, with median values of -2.08, -1.79, -1.55, and -1.19, respectively. SPHs typically display the highest Gini coefficients and the lowest $M_{20}$ values, whereas IRRs show the opposite characteristics. The observed trend from IRRs to SPHs in the figure aligns with our expectations (e.g., \citealt{Zhou+2022, Fang+2023, Dai+2023, Song+2024}).

In Figure \ref{fig:10}, we show the distribution of the $C$ and $G_{2}$ parameters for different types of galaxies.
The parameter $C$ \citep{Conselice+2000, Conselice+2003} is used to describe the concentration of a galaxy's light, with higher $C$ values indicating a more compact light distribution. On the other hand, the $G_2$ parameter \citep{Rosa+2018} reflects the symmetry of gradient vectors in a galaxy's image. Higher $G_2$ values indicate stronger gradient asymmetry, which is common in spiral galaxies, while lower $G_2$ values are typically associated with elliptical galaxies, implying a more symmetric light distribution. In summary, spiral (elliptical) galaxies usually have $G_2$ values greater (less) than 1.
The median values of the $C$ parameter for each galaxy type gradually decrease to 3.59, 3.00, 2.75, and 2.54 as we move from SPHs to IRRs, respectively. The median values of $G_{2}$ gradually increase from SPHs to IRRs to 0.47, 1.53, 1.73, and 1.81, respectively. More compact galaxies tend to exhibit higher $C$ and $G_{2}$ values, whereas those with higher asymmetry display the opposite. Our classification outcomes align with the observed patterns of variation.

The \textit{MID} parameter can be used to measure the irregularity of galaxy morphology \citep{Freeman+2013}. The larger the \textit{MID} value, the more irregular the galaxy's morphology. Conversely, more symmetric galaxies tend to have smaller \textit{MID} values. As illustrated in Figure \ref{fig：11}, the relatively compact and symmetric SPHs and IRRs exhibit smaller \textit{MID} values, whereas the more diffuse and irregular LTD and IRR galaxies show larger \textit{MID} values. The \textit{$\Psi$} parameter describes the clumpy structures in a galaxy's light distribution (\citealt{Law+2007}). From SPHs to IRRs, galaxy concentration decreases, and clumpy structures become more prominent. This observation validates the accuracy of our classification.


\subsubsection{Quantitative Validation of Structural Consistency}
To quantitatively evaluate the reliability of the unsupervised classification results in this study, we reclassified the four morphological classifications (SPH, ETD, LTD, and IRR) obtained by machine learning algorithms into two traditional morphological categories: Early-Type Galaxies (ETGs, synonymous with the SPH class) and Late-Type Galaxies (LTGs, encompassing ETD, LTD, and IRR classes). 
For cross-validation, we utilized two independent structural diagnostic approaches: one based on the S{\'e}rsic index and the other employing the $G$–$M_{20}$ relationship.

The model-based criterion focuses on the S{\'e}rsic index, using the empirical threshold of $n=2.5$ \citep{vander+2008} as the dividing line between galaxies dominated by a bulge and those dominated by a disk. In contrast,  the non-model-based approach employs the $G$–$M_{20}$ classification proposed by \citet{Lotz+2008}.
As depicted in Figure~\ref{fig：12} (a), the distributions of 
ETGs and LTGs in S{\'e}rsic index space show a clear distinction: ETGs with median $n= 4.29$ are primarily found in the region where $n>2.5$, while LTGs with median $n= 1.18$ predominantly reside in the region where $n<2.5$. Utilizing $n=2.5$ as the demarcation, we calculate the ETG probability when $n\ge$2.5, 
$P(n\ge2.5 \mid \mathrm{ETG})$, and the LTG probability when $n<$2.5, $P(n<2.5 \mid \mathrm{LTG}): $
\[
P(n\ge2.5 \mid \mathrm{ETG})=83\%, 
\]
\[
P(n<2.5 \mid \mathrm{LTG})=85\%.
\]
These high probabilities indicate a high level of consistency between the machine learning labels and the classical classification based on luminosity profile morphology within the model-based parameter space. As expected, a larger S{\'e}rsic index corresponds to a luminosity distribution more concentrated towards the center, characteristic of bulge-dominated structures, while a smaller S{\'e}rsic index reflects galaxy systems where disk components are dominant.

Figure~\ref{fig：12} (b) displays the two-dimensional $G$–$M_{20}$ criterion as defined by \citet{Lotz+2008}. The ETG sample is predominantly positioned above the empirical dividing line, characterized by high $G$ and low $M_{20}$ values. Conversely, the LTG sample is mainly found below this line, featuring low $G$ and high $M_{20}$ values. To quantify the dominance of ETGs and LTGs in the individual $G-M_{\rm 20}$ regions to which they belong, we compute the probability of ETGs above the empirical line, $P(G_{\rm high}, M_{\rm 20_{\rm low}} \mid \mathrm{ETG})$, and that of LTGs below the empirical line, $P((G_{\rm low}, M_{\rm 20_{high}}) \mid \mathrm{LTG}):$
\[
P(G_{\rm high}, M_{\rm 20_{\rm low}} \mid \mathrm{ETG}) = 74\%;
\]
\[
P(G_{\rm low}, M_{\rm 20_{high}} \mid \mathrm{LTG}) = 92\%.
\]

The overall completeness rates of the two diagnostic methods are both above 84\%, indicating that the unsupervised morphological classification results are statistically robust, and the resulting classification 
is consistent with the physical characteristics of the galaxies' luminosity structures.

\begin{figure*}
    \includegraphics[width=2\columnwidth]{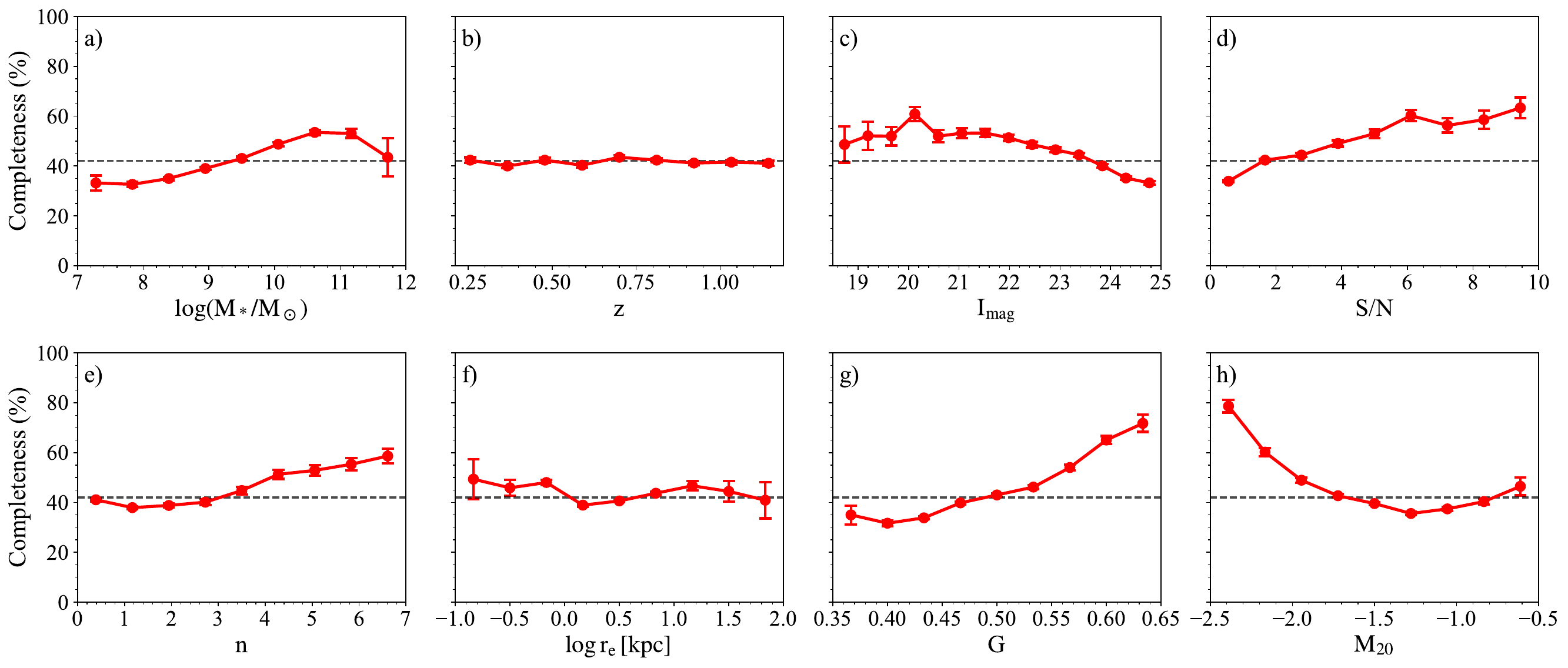}
    \caption{The completeness of the classified sample along various properties, including stellar mass, redshift, magnitude, average signal-to-noise ratio (S/N) per pixel, Sérsic index, effective radius ($r_e$), Gini coefficient ($G$), and $M_{20}$. Completeness is defined as the fraction of high-confidence samples within each parameter bin, with error bars indicating the 95\% confidence interval.}
    \label{fig:13} 
\end{figure*}

\begin{figure*}
\includegraphics[width=2\columnwidth]{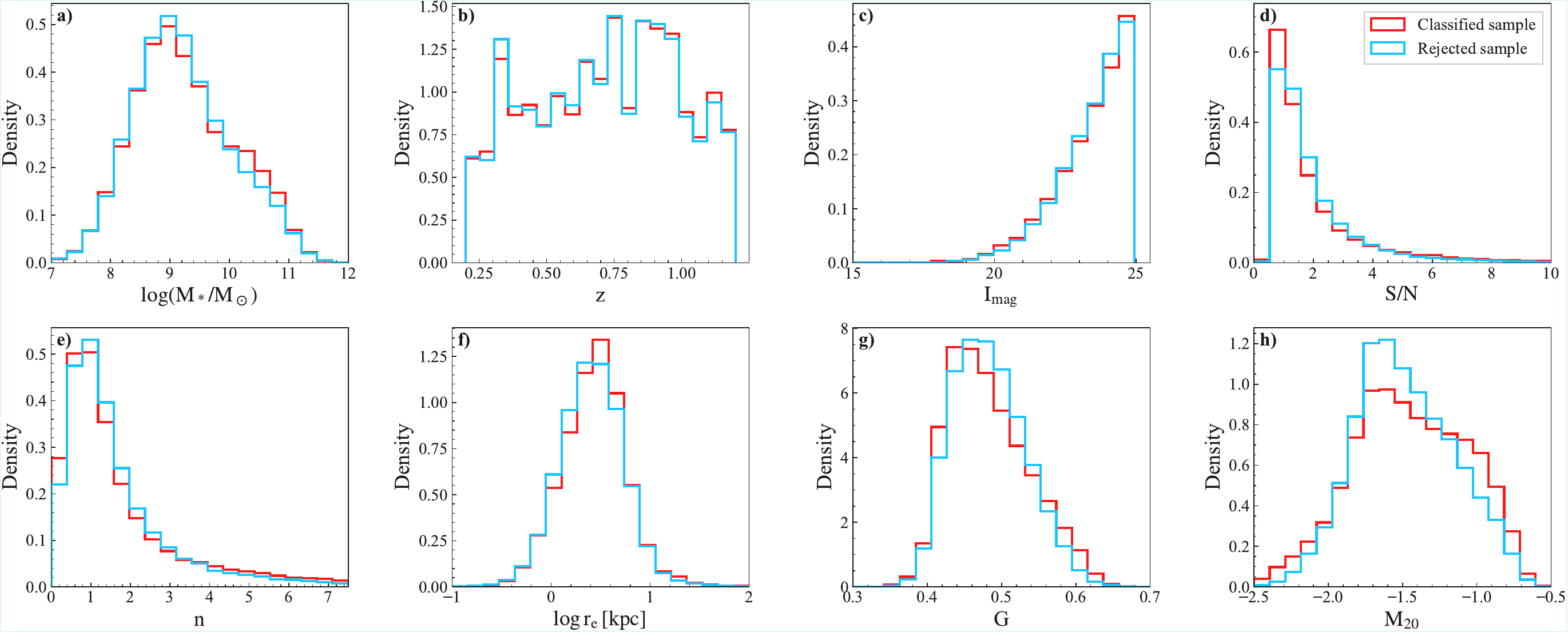}
    \caption{The distribution of Classified sample (red) and rejected sample (blue)  across various physical and structural parameters.}
    \label{fig:14} 
\end{figure*}

\subsection{Evaluation of Catalog Completeness and Scientific Applicability}
To assess the scientific reliability and applicability of the morphological catalog presented in this study, we employed a bagging clustering approach that integrates multiple clustering algorithms. Although this strategy reduced the successfully classified sample size (leaving $\sim$49\% of galaxies rejected), our strict voting mechanism reliably identified a set of galaxies with stable morphologies and high cross-algorithm classification consistency.

To evaluate the catalog's reliability, we analyzed the completeness of the classified sample. Due to their low signal-to-noise ratios (S/Ns) and ambiguous morphological classifications, UNC-type galaxies were excluded from the analysis\footnote{The UNC quasar class is retained and can be used for tasks such as SML to identify galaxies with low S/Ns.}. The resulting high-confidence sample constitutes 42\% of the total. For this sample, we computed completeness curves as a function of various physical and structural parameters, as shown in Figure~\ref{fig:13}. The results indicate that the classification completeness strongly depends on stellar mass, magnitude, signal-to-noise ratio, and certain morphological parameters (such as S{\'e}rsic index $n$, Gini coefficient $G$, and $M_{20}$), while being largely independent of redshift ($z$) and effective radius ($r_e$). Completeness is significantly higher for galaxies with higher S/Ns, brighter magnitudes, and more concentrated structures (i.e., higher $n$ and $G$, lower $M_{20}$). This demonstrates that the bagging-based clustering approach is effective in identifying galaxies with higher resolution and clearer morphological features.

Figure~\ref{fig:14} shows the distributions of successfully classified (51\%) and rejected (49\%) samples across key physical and structural parameters. To quantify the differences between these two groups, we performed Kolmogorov–Smirnov (K–S; \citealt{Kolmogorov+1933}) tests and effect size analyses. The K–S statistic reflects the overall difference between the distributions, Cohen's $d$ (\citealt{Cohen1988}) measures the standardized difference between the means, and Cliff's $\delta$ (\citealt{Cliff+1993}) evaluates the probability of superiority between the groups. For all parameters, the K–S test yields $p$-values below 0.05, while all effect sizes are small (with \(|d| < 0.1\) and \(|\delta| < 0.1\)), indicating that although the differences are statistically significant, the actual discrepancies between successfully classified and rejected samples are minimal.

Notably, the remaining $\sim$49\% of rejected galaxies are retained in the catalog for future analysis and expansion. The currently constructed high-confidence sample can serve as a high-quality training set for supervised, semi-supervised, or transfer learning models, thereby enabling the extension of morphological labels to the entire galaxy population. Additionally, potential differences in parameter distributions between this high-quality training set and the broader target population do not significantly hinder the generalization of morphological learning. As demonstrated in previous studies (e.g., \citealt{DominguezSanchez+2019, walmsley+2021, Song+2024}), models trained on well-defined subsets can be effectively fine-tuned or transferred to noisier and more complex datasets without introducing systematic biases.
\\

\section{Summary} \label{sec:5}
In this work, we improve the initial \texttt{USmorph} framework by introducing a pre-trained ConvNeXt model for feature encoding and combining it with the UMAP dimensionality reduction technique. By optimizing image feature extraction and dimensionality reduction processes, we reduce the number of categories for machine classification from 50 to 20 and save clustering time to one-fifth of the original, significantly decreasing computational resource consumption and enhancing classification visualization efficiency. The improved framework comprises the following four main steps:
\begin{itemize}
    \item  Pre-processing original images using CAE and APCT techniques to reduce noise and enhance rotational invariance. 
    \item Encoding and extracting features from pre-processed data via a pre-trained ConvNeXt model, followed by UMAP dimensionality reduction to further reduce feature dimensions.
    \item Categorizing galaxies into 20 groups using a Bagging-based clustering model. 
\end{itemize}
After visual inspection and removal of sources with inconsistent votes, 51\% (50,056) of galaxies were successfully classified. The visual inspection was then combined to classify 50,056 sources in the COSMOS field into five categories: SPHs, ETDs, LTDs, IRRs, and UNCs. In addition, classification reliability was validated through UMAP visualization analysis and galaxy physical parameter tests. The classification validity is further substantiated by significant differences in appearance characteristics and physical parameters among galaxy types.

Next-generation large-scale sky surveys, including the Legacy Survey of Space and Time at Rubin Observatory and those executed or planned by Euclid, the forthcoming CSST, and the Roman Space Telescope, will produce vast quantities of high-resolution, multi-band images. This necessitates more efficient and adaptable data processing methods. Traditional supervised learning, dependent on annotated data, often experiences reduced recognition accuracy due to domain shift across different sky regions, telescopes, or bands. Our enhanced \texttt{USmorph} algorithm, integrating a ConvNeXt encoder with UMAP dimensionality reduction, significantly cuts computational costs and time for large-scale datasets. Unlike conventional methods, it utilizes pre-trained large models for feature extraction from any dataset, unaffected by variations in observation regions, equipment, or bands. This approach not only maintains high accuracy but also greatly improves the efficiency and adaptability of feature extraction, ideal for automated processing of cross-survey data.

Future work will focus on developing efficient large-model encoding methods and advanced dimensionality reduction techniques to boost machine learning accuracy. Additionally, we aim to explore features of special objects (e.g., tidal tails, gravitational lenses, and little red dots) via subclass subdivision and optimize the algorithm into an intuitive, user-friendly software tool to support efficient technical deployment for survey telescopes on the Chinese space station.

\begin{acknowledgements}
This work is supported by the National Natural Science Foundation of China (NSFC) (grant Nos. 12233008 and 12573012, and Project No. 12503011), the National Key R\&D Program of China (grant No. 2023YFA1608100), the Strategic Priority Research Program of the Chinese Academy of Sciences (grant No. XDB0550200), the Cyrus Chun Ying Tang Foundations, and the 111 Project for “Observational and Theoretical Research on Dark Matter and Dark Energy” (B23042), and the China Manned Space Program with grant No. CMS-CSST-2025-A04. S.L. acknowledges the support from the Key Laboratory of Modern Astronomy and Astrophysics (Nanjing University) by the Ministry of Education. The numerical calculations in this paper have been done on the platform of the High Performance Computing of Anqing Normal University.
\end{acknowledgements}
\bibliography{ref}

@ARTICLE{hubble+1926,
       author = {{Hubble}, E.~P.},
        title = "{Extragalactic nebulae.}",
      journal = {\apj},
         year = 1926,
        month = dec,
       volume = {64},
        pages = {321-369},
          doi = {10.1086/143018},
       adsurl = {https://ui.adsabs.harvard.edu/abs/1926ApJ....64..321H}
}

@INPROCEEDINGS{walmsley+2021,
       author = {{Walmsley}, M.},
        title = "{Galaxy Zoo DECaLS: Detailed Visual Morphology Measurements from Volunteers and Bayesian Deep Learning}",
    booktitle = {American Astronomical Society Meeting Abstracts},
         year = 2021,
       series = {American Astronomical Society Meeting Abstracts},
       volume = {238},
        month = jun,
          eid = {119.02},
        pages = {119.02},
       adsurl = {https://ui.adsabs.harvard.edu/abs/2021AAS...23811902W}
}

@INPROCEEDINGS{Reyes+2018,
  author={Reyes, Esteban and Estévez, Pablo A. and Reyes, Ignacio and Cabrera-Vives, Guillermo and Huijse, Pablo and Carrasco, Rodrigo and Forster, Francisco},
  booktitle={2018 International Joint Conference on Neural Networks (IJCNN)}, 
  title={Enhanced Rotational Invariant Convolutional Neural Network for Supernovae Detection}, 
  year={2018},
  volume={},
  number={},
  pages={1-8},
  doi={10.1109/IJCNN.2018.8489627}}

@ARTICLE{Treu+2023,
       author = {{Treu}, T. and {Calabr{\`o}}, A. and {Castellano}, M. and {Leethochawalit}, N. and {Merlin}, E. and {Fontana}, A. and {Yang}, L. and {Morishita}, T. and {Trenti}, M. and {Dressler}, A. and {Mason}, C. and {Paris}, D. and {Pentericci}, L. and {Roberts-Borsani}, G. and {Vulcani}, B. and {Boyett}, K. and {Bradac}, M. and {Glazebrook}, K. and {Jones}, T. and {Marchesini}, D. and {Mascia}, S. and {Nanayakkara}, T. and {Santini}, P. and {Strait}, V. and {Vanzella}, E. and {Wang}, X.},
        title = "{Early Results From GLASS-JWST. XII. The Morphology of Galaxies at the Epoch of Reionization}",
      journal = {\apjl},
         year = 2023,
        month = jan,
       volume = {942},
       number = {2},
          eid = {L28},
        pages = {L28},
          doi = {10.3847/2041-8213/ac9283},
archivePrefix = {arXiv},
       eprint = {2207.13527},
 primaryClass = {astro-ph.GA},
       adsurl = {https://ui.adsabs.harvard.edu/abs/2023ApJ...942L..28T}
}

@ARTICLE{Ferreira+2023,
       author = {{Ferreira}, Leonardo and {Conselice}, Christopher J. and {Sazonova}, Elizaveta and {Ferrari}, Fabricio and {Caruana}, Joseph and {Tohill}, Cl{\'a}r-Br{\'\i}d and {Lucatelli}, Geferson and {Adams}, Nathan and {Irodotou}, Dimitrios and {Marshall}, Madeline A. and {Roper}, Will J. and {Lovell}, Christopher C. and {Verma}, Aprajita and {Austin}, Duncan and {Trussler}, James and {Wilkins}, Stephen M.},
        title = "{The JWST Hubble Sequence: The Rest-frame Optical Evolution of Galaxy Structure at 1.5 < z < 6.5}",
      journal = {\apj},
         year = 2023,
        month = oct,
       volume = {955},
       number = {2},
          eid = {94},
        pages = {94},
          doi = {10.3847/1538-4357/acec76},
archivePrefix = {arXiv},
       eprint = {2210.01110},
 primaryClass = {astro-ph.GA},
       adsurl = {https://ui.adsabs.harvard.edu/abs/2023ApJ...955...94F}
}

@misc{lang+2016,
  author = {Lang, Dustin and Hogg, David W. and Mykytyn, David},
  title = {The Tractor: Probabilistic astronomical source detection and measurement},
  howpublished = {Astrophysics Source Code Library, record ascl:1604.008},
  year = 2016,
  month = apr,
  eid = {ascl:1604.008},
  url = {https://ui.adsabs.harvard.edu/abs/2016ascl.soft04008L},
  note = {Provided by the SAO/NASA Astrophysics Data System}
}

@article{Kolesnikov+2023,
    author = {Kolesnikov, I and Sampaio, V M and de Carvalho, R R and Conselice, C and Rembold, S B and Mendes, C L and Rosa, R R},
    title = {Unveiling galaxy morphology through an unsupervised-supervised hybrid approach},
    journal = {Monthly Notices of the Royal Astronomical Society},
    volume = {528},
    number = {1},
    pages = {82-107},
    year = {2023},
    month = {12},
    issn = {0035-8711},
    doi = {10.1093/mnras/stad3934},
    url = {https://doi.org/10.1093/mnras/stad3934},
    eprint = {https://academic.oup.com/mnras/article-pdf/528/1/82/55675427/stad3934.pdf},
}

@ARTICLE{Conselice+2003,
       author = {{Conselice}, Christopher J.},
        title = "{The Relationship between Stellar Light Distributions of Galaxies and Their Formation Histories}",
      journal = {\apjs},
         year = 2003,
        month = jul,
       volume = {147},
       number = {1},
        pages = {1-28},
          doi = {10.1086/375001},
archivePrefix = {arXiv},
       eprint = {astro-ph/0303065},
 primaryClass = {astro-ph},
       adsurl = {https://ui.adsabs.harvard.edu/abs/2003ApJS..147....1C}
}

@inproceedings{Chen+2020,
  author = {Chen, Ting and Kornblith, Simon and Norouzi, Mohammad and Hinton, Geoffrey},
  title = {A simple framework for contrastive learning of visual representations},
  booktitle = {Proceedings of the 37th International Conference on Machine Learning}, 
  year = {2020},
  publisher = {JMLR.org},
  articleno = {149},
  numpages = {11},
  series = {ICML'20}
}

@ARTICLE{Rosa+2018,
       author = {{Rosa}, R.~R. and {de Carvalho}, R.~R. and {Sautter}, R.~A. and {Barchi}, P.~H. and {Stalder}, D.~H. and {Moura}, T.~C. and {Rembold}, S.~B. and {Morell}, D.~R.~F. and {Ferreira}, N.~C.},
        title = "{Gradient pattern analysis applied to galaxy morphology}",
      journal = {\mnras},
         year = 2018,
        month = jun,
       volume = {477},
       number = {1},
        pages = {L101-L105},
          doi = {10.1093/mnrasl/sly054},
archivePrefix = {arXiv},
       eprint = {1803.10853},
 primaryClass = {astro-ph.GA},
       adsurl = {https://ui.adsabs.harvard.edu/abs/2018MNRAS.477L.101R}
}

@article{gong+2018,
  title={On the Intrinsic Dimensionality of Image Representations},
  author={Sixue Gong and Vishnu Naresh Boddeti and Anil K. Jain},
  journal={2019 IEEE/CVF Conference on Computer Vision and Pattern Recognition (CVPR)},
  year={2018},
  pages={3982-3991},
  url={https://api.semanticscholar.org/CorpusID:125095565}
}

@article{Mao+2024,
author = {Jialin Mao  and Itay Griniasty  and Han Kheng Teoh  and Rahul Ramesh  and Rubing Yang  and Mark K. Transtrum  and James P. Sethna  and Pratik Chaudhari },
title = {The training process of many deep networks explores the same low-dimensional manifold},
journal = {Proceedings of the National Academy of Sciences},
volume = {121},
number = {12},
pages = {e2310002121},
year = {2024},
doi = {10.1073/pnas.2310002121},
URL = {https://www.pnas.org/doi/abs/10.1073/pnas.2310002121},
eprint = {https://www.pnas.org/doi/pdf/10.1073/pnas.2310002121}
}

@inproceedings{lowe+2024,
  author    = {Lowe, S. C. and Haurum, J. B. and Oore, S. and Moeslund, T. B. and Taylor, G. W.},
  title     = {An Empirical Study into Clustering of Unseen Datasets with Self-Supervised Foundation Models},
  booktitle = {ICML 2024 Workshop on Foundation Models in the Wild},
  year      = {2024}
}

@article{Jang+2024,
    author = {Jang, Hojin and Tong, Frank},
    title = {Improved modeling of human vision by incorporating robustness to blur in convolutional neural networks},
    journal = {Nature Communications},
    year = {2024},
    volume = {15},
    number = {1},
    pages = {1989},
    month = {mar},
    issn = {2041-1723},
    doi = {10.1038/s41467-024-45679-0},
    url = {https://doi.org/10.1038/s41467-024-45679-0}
}

@misc{dosovitskiy+2021,
      title={An Image is Worth 16x16 Words: Transformers for Image Recognition at Scale}, 
      author={Alexey Dosovitskiy and Lucas Beyer and Alexander Kolesnikov and Dirk Weissenborn and Xiaohua Zhai and Thomas Unterthiner and Mostafa Dehghani and Matthias Minderer and Georg Heigold and Sylvain Gelly and Jakob Uszkoreit and Neil Houlsby},
      year={2021},
      eprint={2010.11929},
      archivePrefix={arXiv},
      primaryClass={cs.CV},
      url={https://arxiv.org/abs/2010.11929}, 
}

@INPROCEEDINGS{Woo+2023,
  author={Woo, Sanghyun and Debnath, Shoubhik and Hu, Ronghang and Chen, Xinlei and Liu, Zhuang and Kweon, In So and Xie, Saining},
  booktitle={2023 IEEE/CVF Conference on Computer Vision and Pattern Recognition (CVPR)}, 
  title={ConvNeXt V2: Co-designing and Scaling ConvNets with Masked Autoencoders}, 
  year={2023},
  volume={},
  number={},
  pages={16133-16142},
  doi={10.1109/CVPR52729.2023.01548}}

@ARTICLE{Stoughton+2002,
       author = {{Stoughton}, Chris and {Lupton}, Robert H. and {Bernardi}, Mariangela and {Blanton}, Michael R. and {Burles}, Scott and {Castander}, Francisco J. and {Connolly}, A.~J. and {Eisenstein}, Daniel J. and {Frieman}, Joshua A. and {Hennessy}, G.~S. and {Hindsley}, Robert B. and {Ivezi{\'c}}, {\v{Z}}eljko and {Kent}, Stephen and {Kunszt}, Peter Z. and {Lee}, Brian C. and {Meiksin}, Avery and {Munn}, Jeffrey A. and {Newberg}, Heidi Jo and {Nichol}, R.~C. and {Nicinski}, Tom and {Pier}, Jeffrey R. and {Richards}, Gordon T. and {Richmond}, Michael W. and {Schlegel}, David J. and {Smith}, J. Allyn and {Strauss}, Michael A. and {SubbaRao}, Mark and {Szalay}, Alexander S. and {Thakar}, Aniruddha R. and {Tucker}, Douglas L. and {Vanden Berk}, Daniel E. and {Yanny}, Brian and {Adelman}, Jennifer K. and {Anderson}, Jr., John E. and {Anderson}, Scott F. and {Annis}, James and {Bahcall}, Neta A. and {Bakken}, J.~A. and {Bartelmann}, Matthias and {Bastian}, Steven and {Bauer}, Amanda and {Berman}, Eileen and {B{\"o}hringer}, Hans and {Boroski}, William N. and {Bracker}, Steve and {Briegel}, Charlie and {Briggs}, John W. and {Brinkmann}, J. and {Brunner}, Robert and {Carey}, Larry and {Carr}, Michael A. and {Chen}, Bing and {Christian}, Damian and {Colestock}, Patrick L. and {Crocker}, J.~H. and {Csabai}, Istv{\'a}n and {Czarapata}, Paul C. and {Dalcanton}, Julianne and {Davidsen}, Arthur F. and {Davis}, John Eric and {Dehnen}, Walter and {Dodelson}, Scott and {Doi}, Mamoru and {Dombeck}, Tom and {Donahue}, Megan and {Ellman}, Nancy and {Elms}, Brian R. and {Evans}, Michael L. and {Eyer}, Laurent and {Fan}, Xiaohui and {Federwitz}, Glenn R. and {Friedman}, Scott and {Fukugita}, Masataka and {Gal}, Roy and {Gillespie}, Bruce and {Glazebrook}, Karl and {Gray}, Jim and {Grebel}, Eva K. and {Greenawalt}, Bruce and {Greene}, Gretchen and {Gunn}, James E. and {de Haas}, Ernst and {Haiman}, Zolt{\'a}n and {Haldeman}, Merle and {Hall}, Patrick B. and {Hamabe}, Masaru and {Hansen}, Brad and {Harris}, Frederick H. and {Harris}, Hugh and {Harvanek}, Michael and {Hawley}, Suzanne L. and {Hayes}, J.~J.~E. and {Heckman}, Timothy M. and {Helmi}, Amina and {Henden}, Arne and {Hogan}, Craig J. and {Hogg}, David W. and {Holmgren}, Donald J. and {Holtzman}, Jon and {Huang}, Chih-Hao and {Hull}, Charles and {Ichikawa}, Shin-Ichi and {Ichikawa}, Takashi and {Johnston}, David E. and {Kauffmann}, Guinevere and {Kim}, Rita S.~J. and {Kimball}, Tim and {Kinney}, E. and {Klaene}, Mark and {Kleinman}, S.~J. and {Klypin}, Anatoly and {Knapp}, G.~R. and {Korienek}, John and {Krolik}, Julian and {Kron}, Richard G. and {Krzesi{\'n}ski}, Jurek and {Lamb}, D.~Q. and {Leger}, R. French and {Limmongkol}, Siriluk and {Lindenmeyer}, Carl and {Long}, Daniel C. and {Loomis}, Craig and {Loveday}, Jon and {MacKinnon}, Bryan and {Mannery}, Edward J. and {Mantsch}, P.~M. and {Margon}, Bruce and {McGehee}, Peregrine and {McKay}, Timothy A. and {McLean}, Brian and {Menou}, Kristen and {Merelli}, Aronne and {Mo}, H.~J. and {Monet}, David G. and {Nakamura}, Osamu and {Narayanan}, Vijay K. and {Nash}, Thomas and {Neilsen}, Jr., Eric H. and {Newman}, Peter R. and {Nitta}, Atsuko and {Odenkirchen}, Michael and {Okada}, Norio and {Okamura}, Sadanori and {Ostriker}, Jeremiah P. and {Owen}, Russell and {Pauls}, A. George and {Peoples}, John and {Peterson}, R.~S. and {Petravick}, Donald and {Pope}, Adrian and {Pordes}, Ruth and {Postman}, Marc and {Prosapio}, Angela and {Quinn}, Thomas R. and {Rechenmacher}, Ron and {Rivetta}, Claudio H. and {Rix}, Hans-Walter and {Rockosi}, Constance M. and {Rosner}, Robert and {Ruthmansdorfer}, Kurt and {Sandford}, Dale and {Schneider}, Donald P. and {Scranton}, Ryan and {Sekiguchi}, Maki and {Sergey}, Gary and {Sheth}, Ravi and {Shimasaku}, Kazuhiro and {Smee}, Stephen and {Snedden}, Stephanie A. and {Stebbins}, Albert and {Stubbs}, Christopher and {Szapudi}, Istv{\'a}n and {Szkody}, Paula and {Szokoly}, Gyula P. and {Tabachnik}, Serge and {Tsvetanov}, Zlatan and {Uomoto}, Alan and {Vogeley}, Michael S. and {Voges}, Wolfgang and {Waddell}, Patrick and {Walterbos}, Ren{\'e} and {Wang}, Shu-i. and {Watanabe}, Masaru and {Weinberg}, David H. and {White}, Richard L. and {White}, Simon D.~M. and {Wilhite}, Brian and {Wolfe}, David and {Yasuda}, Naoki and {York}, Donald G. and {Zehavi}, Idit and {Zheng}, Wei},
        title = "{Sloan Digital Sky Survey: Early Data Release}",
      journal = {\aj},
         year = 2002,
        month = jan,
       volume = {123},
       number = {1},
        pages = {485-548},
          doi = {10.1086/324741},
       adsurl = {https://ui.adsabs.harvard.edu/abs/2002AJ....123..485S}
}

@article{McInnes+2018,
  title={UMAP: Uniform Manifold Approximation and Projection for Dimension Reduction},
  author={Leland McInnes and John Healy},
  journal={ArXiv},
  year={2018},
  volume={abs/1802.03426},
  url={https://api.semanticscholar.org/CorpusID:3641284}
}

@article{Law+2007,
doi = {10.1086/510357},
url = {https://dx.doi.org/10.1086/510357},
year = {2007},
month = {feb},
publisher = {},
volume = {656},
number = {1},
pages = {1},
author = {David R. Law and Charles C. Steidel and Dawn K. Erb and Max Pettini and Naveen A. Reddy and Alice E. Shapley and Kurt L. Adelberger and David J. Simenc},
title = {The Physical Nature of Rest-UV Galaxy Morphology during the Peak Epoch of Galaxy Formation},
journal = {The Astrophysical Journal}
}

@article{Weaver+2022,
  author = {{Weaver}, J.~R. and {Kauffmann}, O.~B. and {Ilbert}, O. and {McCracken}, H.~J. and {Moneti}, A. and {Toft}, S. and {Brammer}, G. and {Shuntov}, M. and {Davidzon}, I. and {Hsieh}, B.~C. and {Laigle}, C. and {Anastasiou}, A. and {Jespersen}, C.~K. and {Vinther}, J. and {Capak}, P. and {Casey}, C.~M. and {McPartland}, C.~J.~R. and {Milvang-Jensen}, B. and {Mobasher}, B. and {Sanders}, D.~B. and {Zalesky}, L. and {Arnouts}, S. and {Aussel}, H. and {Dunlop}, J.~S. and {Faisst}, A. and {Franx}, M. and {Furtak}, L.~J. and {Fynbo}, J.~P.~U. and {Gould}, K.~M.~L. and {Greve}, T.~R. and {Gwyn}, S. and {Kartaltepe}, J.~S. and {Kashino}, D. and {Koekemoer}, A.~M. and {Kokorev}, V. and {Le Fèvre}, O. and {Lilly}, S. and {Masters}, D. and {Magdis}, G. and {Mehta}, V. and {Peng}, Y. and {Riechers}, D.~A. and {Salvato}, M. and {Sawicki}, M. and {Scarlata}, C. and {Scoville}, N. and {Shirley}, R. and {Silverman}, J.~D. and {Sneppen}, A. and {Smolčić}, V. and {Steinhardt}, C. and {Stern}, D. and {Tanaka}, M. and {Taniguchi}, Y. and {Teplitz}, H.~I. and {Vaccari}, M. and {Wang}, W.-H. and {Zamorani}, G.},
  title = {COSMOS2020: A Panchromatic View of the Universe to z ∼ 10 from Two Complementary Catalogs},
  journal = {The Astrophysical Journal Supplement Series},
  year = {2022},
  month = {jan},
  volume = {258},
  number = {1},
  pages = {11},
  doi = {10.3847/1538-4365/ac3078},
  url = {https://dx.doi.org/10.3847/1538-4365/ac3078}
}

@ARTICLE{Oke+1983,
       author = {{Oke}, J.~B. and {Gunn}, J.~E.},
        title = "{Secondary standard stars for absolute spectrophotometry.}",
      journal = {\apj},
         year = 1983,
        month = mar,
       volume = {266},
        pages = {713-717},
          doi = {10.1086/160817},
       adsurl = {https://ui.adsabs.harvard.edu/abs/1983ApJ...266..713O}
}

@article{Chabrier+2003,
doi = {10.1086/376392},
url = {https://dx.doi.org/10.1086/376392},
year = {2003},
month = {jul},
publisher = {The University of Chicago Press},
volume = {115},
number = {809},
pages = {763},
author = {Gilles Chabrier},
title = {Galactic Stellar and Substellar Initial Mass Function1},
journal = {Publications of the Astronomical Society of the Pacific},
}

@article{Kauffmann+2004,
  title = {The Environmental Dependence of the Relations between Stellar Mass, Structure, Star Formation and Nuclear Activity in Galaxies: {{Galaxy}} Structure, Star Formation and Nuclear Activity},
  shorttitle = {The Environmental Dependence of the Relations between Stellar Mass, Structure, Star Formation and Nuclear Activity in Galaxies},
  author = {Kauffmann, Guinevere and White, Simon D. M. and Heckman, Timothy M. and M{\'e}nard, Brice and Brinchmann, Jarle and Charlot, St{\'e}phane and Tremonti, Christy and Brinkmann, Jon},
  year = {2004},
  journal = {Monthly Notices of the Royal Astronomical Society},
  volume = {353},
  number = {3},
  pages = {713--731},
  doi = {10.1111/j.1365-2966.2004.08117.x},
  urldate = {2022-08-26},
  langid = {english}
}

@article{omand+2014,
  title = {The Connection between Galaxy Structure and Quenching Efficiency},
  author = {Omand, Conor M. B. and Balogh, Michael L. and Poggianti, Bianca M.},
  year = {2014},
  journal = {Monthly Notices of the Royal Astronomical Society},
  volume = {440},
  number = {1},
  pages = {843--858},
  doi = {10.1093/mnras/stu331},
  urldate = {2023-07-11},
  langid = {english}
}

@ARTICLE{su+2025,
       author = {{Su}, Yuguo and {Fang}, Guanwen and {Lu}, Shiying and {Lin}, Zesen},
        title = "{The impact of morphological quenching mechanisms on star formation activity at 0.2 < z < 1.2 in the COSMOS field}",
      journal = {\aap},
         year = 2025,
        month = jul,
       volume = {699},
          eid = {A184},
        pages = {A184},
          doi = {10.1051/0004-6361/202553693},
       adsurl = {https://ui.adsabs.harvard.edu/abs/2025A&A...699A.184S}
}

@article{simmons+2014,
    author = {Simmons, B. D. and Melvin, Thomas and Lintott, Chris and Masters, Karen L. and Willett, Kyle W. and Keel, William C. and Smethurst, R. J. and Cheung, Edmond and Nichol, Robert C. and Schawinski, Kevin and Rutkowski, Michael and Kartaltepe, Jeyhan S. and Bell, Eric F. and Casteels, Kevin R. V. and Conselice, Christopher J. and Almaini, Omar and Ferguson, Henry C. and Fortson, Lucy and Hartley, William and Kocevski, Dale and Koekemoer, Anton M. and McIntosh, Daniel H. and Mortlock, Alice and Newman, Jeffrey A. and Ownsworth, Jamie and Bamford, Steven and Dahlen, Tomas and Faber, Sandra M. and Finkelstein, Steven L. and Fontana, Adriano and Galametz, Audrey and Grogin, N. A. and Grützbauch, Ruth and Guo, Yicheng and Häußler, Boris and Jek, Kian J. and Kaviraj, Sugata and Lucas, Ray A. and Peth, Michael and Salvato, Mara and Wiklind, Tommy and Wuyts, Stijn},
    title = {Galaxy Zoo: CANDELS barred discs and bar fractions},
    journal = {Monthly Notices of the Royal Astronomical Society},
    volume = {445},
    number = {4},
    pages = {3466-3474},
    year = {2014},
    month = {10},
    issn = {0035-8711},
    doi = {10.1093/mnras/stu1817},
    url = {https://doi.org/10.1093/mnras/stu1817},
    eprint = {https://academic.oup.com/mnras/article-pdf/445/4/3466/6076511/stu1817.pdf}
}

@article{kawinwanichakij+2017,
  title = {Effect of {{Local Environment}} and {{Stellar Mass}} on {{Galaxy Quenching}} and {{Morphology}} at 0.5 {$<$} z {$<$} 2.0},
  author = {Kawinwanichakij, Lalitwadee and Papovich, Casey and Quadri, Ryan F. and Glazebrook, Karl and Kacprzak, Glenn G. and Allen, Rebecca J. and Bell, Eric F. and Croton, Darren J. and Dekel, Avishai and Ferguson, Henry C. and Forrest, Ben and Grogin, Norman A. and Guo, Yicheng and Kocevski, Dale D. and Koekemoer, Anton M. and Labb{\'e}, Ivo and Lucas, Ray A. and Nanayakkara, Themiya and Spitler, Lee R. and Straatman, Caroline M. S. and Tran, Kim-Vy H. and Tomczak, Adam and van Dokkum, Pieter},
  year = {2017},
  journal = {The Astrophysical Journal},
  volume = {847},
  number = {2},
  pages = {134},
  doi = {10.3847/1538-4357/aa8b75},
  urldate = {2021-12-06},
  langid = {english}
}

@ARTICLE{flugge+1959,
       author = {{Fl{\"u}gge}, S.},
        title = "{Astrophysik IV: Sternsysteme / Astrophysics IV: Stellar Systems}",
      journal = {Handbuch der Physik},
         year = 1959,
        month = jan,
       volume = {11},
          doi = {10.1007/978-3-642-45932-0},
       adsurl = {https://ui.adsabs.harvard.edu/abs/1959HDP....53.....F}
}

@INPROCEEDINGS{yao+2019,
  author={Yao, Xiwen and Feng, Xiaoxu and Cheng, Gong and Han, Junwei and Guo, Lei},
  booktitle={IGARSS 2019 - 2019 IEEE International Geoscience and Remote Sensing Symposium}, 
  title={Rotation-Invariant Latent Semantic Representation Learning for Object Detection in VHR Optical Remote Sensing Images}, 
  year={2019},
  volume={},
  number={},
  pages={1382-1385},
  doi={10.1109/IGARSS.2019.8899285}}

@article{Scoville+2007,
   title={The Cosmic Evolution Survey (COSMOS): Overview},
   volume={172},
   ISSN={1538-4365},
   url={http://dx.doi.org/10.1086/516585},
   DOI={10.1086/516585},
   number={1},
   journal={The Astrophysical Journal Supplement Series},
   publisher={American Astronomical Society},
   author={Scoville, N. and Aussel, H. and Brusa, M. and Capak, P. and Carollo, C. M. and Elvis, M. and Giavalisco, M. and Guzzo, L. and Hasinger, G. and Impey, C. and Kneib, J.‐P. and LeFevre, O. and Lilly, S. J. and Mobasher, B. and Renzini, A. and Rich, R. M. and Sanders, D. B. and Schinnerer, E. and Schminovich, D. and Shopbell, P. and Taniguchi, Y. and Tyson, N. D.},
   year={2007},
   month=sep, pages={1–8} }

@article{Dieleman+2015,
   title={Rotation-invariant convolutional neural networks for galaxy morphology prediction},
   volume={450},
   ISSN={0035-8711},
   url={http://dx.doi.org/10.1093/mnras/stv632},
   DOI={10.1093/mnras/stv632},
   number={2},
   journal={Monthly Notices of the Royal Astronomical Society},
   publisher={Oxford University Press (OUP)},
   author={Dieleman, Sander and Willett, Kyle W. and Dambre, Joni},
   year={2015},
   month=apr, pages={1441–1459} }

@INPROCEEDINGS{Fernando+2024,
  author={Fernando, Thrinith and Rathnayake, Samadhi and Dissanayaka, Kapila},
  booktitle={2024 6th International Conference on Advancements in Computing (ICAC)}, 
  title={Galaxy Morphology Classification Based on VGG19 Deep Convolutional Neural Network}, 
  year={2024},
  volume={},
  number={},
  pages={229-234},
  doi={10.1109/ICAC64487.2024.10851157}}

@article{York+2000,
doi = {10.1086/301513},
url = {https://dx.doi.org/10.1086/301513},
year = {2000},
month = {sep},
publisher = {},
volume = {120},
number = {3},
pages = {1579},
author = {York, Donald G. and Adelman, J. and Anderson, Jr., John E. and Anderson, Scott F. and Annis, James and Bahcall, Neta A. and Bakken, J. A. and Barkhouser, Robert and Bastian, Steven and Berman, Eileen and Boroski, William N. and Bracker, Steve and Briegel, Charlie and Briggs, John W. and Brinkmann, J. and Brunner, Robert and Burles, Scott and Carey, Larry and Carr, Michael A. and Castander, Francisco J. and Chen, Bing and Colestock, Patrick L. and Connolly, A. J. and Crocker, J. H. and Csabai, István and Czarapata, Paul C. and Davis, John Eric and Doi, Mamoru and Dombeck, Tom and Eisenstein, Daniel and Ellman, Nancy and Elms, Brian R. and Evans, Michael L. and Fan, Xiaohui and Federwitz, Glenn R. and Fiscelli, Larry and Friedman, Scott and Frieman, Joshua A. and Fukugita, Masataka and Gillespie, Bruce and Gunn, James E. and Gurbani, Vijay K. and de Haas, Ernst and Haldeman, Merle and Harris, Frederick H. and Hayes, J. and Heckman, Timothy M. and Hennessy, G. S. and Hindsley, Robert B. and Holm, Scott and Holmgren, Donald J. and Huang, Chi-hao and Hull, Charles and Husby, Don and Ichikawa, Shin-Ichi and Ichikawa, Takashi and Ivezić, Željko and Kent, Stephen and Kim, Rita S. J. and Kinney, E. and Klaene, Mark and Kleinman, A. N. and Kleinman, S. and Knapp, G. R. and Korienek, John and Kron, Richard G. and Kunszt, Peter Z. and Lamb, D. Q. and Lee, B. and Leger, R. French and Limmongkol, Siriluk and Lindenmeyer, Carl and Long, Daniel C. and Loomis, Craig and Loveday, Jon and Lucinio, Rich and Lupton, Robert H. and MacKinnon, Bryan and Mannery, Edward J. and Mantsch, P. M. and Margon, Bruce and McGehee, Peregrine and McKay, Timothy A. and Meiksin, Avery and Merelli, Aronne and Monet, David G. and Munn, Jeffrey A. and Narayanan, Vijay K. and Nash, Thomas and Neilsen, Eric and Neswold, Rich and Newberg, Heidi Jo and Nichol, R. C. and Nicinski, Tom and Nonino, Mario and Okada, Norio and Okamura, Sadanori and Ostriker, Jeremiah P. and Owen, Russell and Pauls, A. George and Peoples, John and Peterson, R. L. and Petravick, Donald and Pier, Jeffrey R. and Pope, Adrian and Pordes, Ruth and Prosapio, Angela and Rechenmacher, Ron and Quinn, Thomas R. and Richards, Gordon T. and Richmond, Michael W. and Rivetta, Claudio H. and Rockosi, Constance M. and Ruthmansdorfer, Kurt and Sandford, Dale and Schlegel, David J. and Schneider, Donald P. and Sekiguchi, Maki and Sergey, Gary and Shimasaku, Kazuhiro and Siegmund, Walter A. and Smee, Stephen and Smith, J. Allyn and Snedden, S. and Stone, R. and Stoughton, Chris and Strauss, Michael A. and Stubbs, Christopher and SubbaRao, Mark and Szalay, Alexander S. and Szapudi, Istvan and Szokoly, Gyula P. and Thakar, Anirudda R. and Tremonti, Christy and Tucker, Douglas L. and Uomoto, Alan and Vanden Berk, Dan and Vogeley, Michael S. and Waddell, Patrick and Wang, Shu-i and Watanabe, Masaru and Weinberg, David H. and Yanny, Brian and Yasuda, Naoki},
title = {The Sloan Digital Sky Survey: Technical Summary},
journal = {The Astronomical Journal}
}

@article{Zhou+2022,
   title={Automatic Morphological Classification of Galaxies: Convolutional Autoencoder and Bagging-based Multiclustering Model},
   volume={163},
   ISSN={1538-3881},
   url={http://dx.doi.org/10.3847/1538-3881/ac4245},
   DOI={10.3847/1538-3881/ac4245},
   number={2},
   journal={The Astronomical Journal},
   publisher={American Astronomical Society},
   author={Zhou, ChiChun and Gu, Yizhou and Fang, Guanwen and Lin, Zesen},
   year={2022},
   month=jan, pages={86} }

@INPROCEEDINGS{liu+2022,
  author={Liu, Zhuang and Mao, Hanzi and Wu, Chao-Yuan and Feichtenhofer, Christoph and Darrell, Trevor and Xie, Saining},
  booktitle={2022 IEEE/CVF Conference on Computer Vision and Pattern Recognition (CVPR)}, 
  title={A ConvNet for the 2020s}, 
  year={2022},
  volume={},
  number={},
  pages={11966-11976},
  doi={10.1109/CVPR52688.2022.01167}}

@article{Grill+2020,
  title={Bootstrap Your Own Latent: A New Approach to Self-Supervised Learning},
  author={Jean-Bastien Grill and Florian Strub and Florent Altch'e and Corentin Tallec and Pierre H. Richemond and Elena Buchatskaya and Carl Doersch and Bernardo {\'A}vila Pires and Zhaohan Daniel Guo and Mohammad Gheshlaghi Azar and Bilal Piot and Koray Kavukcuoglu and R{\'e}mi Munos and Michal Valko},
  journal={ArXiv},
  year={2020},
  volume={abs/2006.07733},
  url={https://api.semanticscholar.org/CorpusID:219687798}
}

@misc{he2+2015,
      title={Deep Residual Learning for Image Recognition}, 
      author={Kaiming He and Xiangyu Zhang and Shaoqing Ren and Jian Sun},
      year={2015},
      eprint={1512.03385},
      archivePrefix={arXiv},
      primaryClass={cs.CV}
}

@article{Fang+2023,
doi = {10.3847/1538-3881/aca1a6},
url = {https://dx.doi.org/10.3847/1538-3881/aca1a6},
year = {2023},
month = {jan},
publisher = {The American Astronomical Society},
volume = {165},
number = {2},
pages = {35},
author = {GuanWen Fang and Shuo Ba and Yizhou Gu and Zesen Lin and Yuejie Hou and Chenxin Qin and Chichun Zhou and Jun Xu and Yao Dai and Jie Song and Xu Kong},
title = {Automatic Classification of Galaxy Morphology: A Rotationally-invariant Supervised Machine-learning Method Based on the Unsupervised Machine-learning Data Set},
journal = {The Astronomical Journal}
}

@article{van+2008,
title = "Visualizing High-Dimensional Data Using t-SNE",
author = "{van der Maaten}, L.J.P. and G.E. Hinton",
note = "Pagination: 27",
year = "2008",
language = "English",
volume = "9",
pages = "2579--2605",
journal = "Journal of Machine Learning Research",
issn = "1532-4435",
publisher = "Microtome Publishing",
number = "nov",
}

@article{Bramme+2008,
doi = {10.1086/591786},
url = {https://dx.doi.org/10.1086/591786},
year = {2008},
month = {oct},
publisher = {},
volume = {686},
number = {2},
pages = {1503},
author = {Gabriel B. Brammer and Pieter G. van Dokkum and Paolo Coppi},
title = {EAZY: A Fast, Public Photometric Redshift Code},
journal = {The Astrophysical Journal}
}

@article{Ilbert+2009,
doi = {10.1088/0004-637X/690/2/1236},
url = {https://dx.doi.org/10.1088/0004-637X/690/2/1236},
year = {2008},
month = {dec},
publisher = {The American Astronomical Society},
volume = {690},
number = {2},
pages = {1236},
author = {O. Ilbert and P. Capak and M. Salvato and H. Aussel and H. J. McCracken and D. B. Sanders and N. Scoville and J. Kartaltepe and S. Arnouts and E. Le Floc'h and B. Mobasher and Y. Taniguchi and F. Lamareille and A. Leauthaud and S. Sasaki and D. Thompson and M. Zamojski and G. Zamorani and S. Bardelli and M. Bolzonella and A. Bongiorno and M. Brusa and K. I. Caputi and C. M. Carollo and T. Contini and R. Cook and G. Coppa and O. Cucciati and S. de la Torre and L. de Ravel and P. Franzetti and B. Garilli and G. Hasinger and A. Iovino and P. Kampczyk and J.-P. Kneib and C. Knobel and K. Kovac and J. F. Le Borgne and V. Le Brun and O. Le Fèvre and S. Lilly and D. Looper and C. Maier and V. Mainieri and Y. Mellier and M. Mignoli and T. Murayama and R. Pellò and Y. Peng and E. Pérez-Montero and A. Renzini and E. Ricciardelli and D. Schiminovich and M. Scodeggio and Y. Shioya and J. Silverman and J. Surace and M. Tanaka and L. Tasca and L. Tresse and D. Vergani and E. Zucca},
title = {COSMOS PHOTOMETRIC REDSHIFTS WITH 30-BANDS FOR 2-deg2},
journal = {The Astrophysical Journal}
}

@article{Peng+2002,
doi = {10.1086/340952},
url = {https://dx.doi.org/10.1086/340952},
year = {2002},
month = {jul},
publisher = {},
volume = {124},
number = {1},
pages = {266},
author = {Chien Y. Peng and Luis C. Ho and Chris D. Impey and Hans-Walter Rix},
title = {Detailed Structural Decomposition of Galaxy
Images*},
journal = {The Astronomical Journal}
}

@article{Lotz+2004,
doi = {10.1086/421849},
url = {https://dx.doi.org/10.1086/421849},
year = {2004},
month = {jul},
publisher = {},
volume = {128},
number = {1},
pages = {163},
author = {Jennifer M. Lotz and Joel Primack and Piero Madau},
title = {A New Nonparametric Approach to Galaxy Morphological Classification},
journal = {The Astronomical Journal}
}

@ARTICLE{vander+2008,
       author = {{van der Wel}, Arjen and {Holden}, Bradford P. and {Zirm}, Andrew W. and {Franx}, Marijn and {Rettura}, Alessandro and {Illingworth}, Garth D. and {Ford}, Holland C.},
        title = "{Recent Structural Evolution of Early-Type Galaxies: Size Growth from z = 1 to z = 0}",
      journal = {\apj},
         year = 2008,
        month = nov,
       volume = {688},
       number = {1},
        pages = {48-58},
          doi = {10.1086/592267},
archivePrefix = {arXiv},
       eprint = {0808.0077},
 primaryClass = {astro-ph},
       adsurl = {https://ui.adsabs.harvard.edu/abs/2008ApJ...688...48V}
}

@ARTICLE{Lecun+1998,
  author={Lecun, Y. and Bottou, L. and Bengio, Y. and Haffner, P.},
  journal={Proceedings of the IEEE}, 
  title={Gradient-based learning applied to document recognition}, 
  year={1998},
  volume={86},
  number={11},
  pages={2278-2324},
  doi={10.1109/5.726791}}

@article{Davis+2014,
doi = {10.1088/0004-637X/790/2/87},
url = {https://dx.doi.org/10.1088/0004-637X/790/2/87},
year = {2014},
month = {jul},
publisher = {The American Astronomical Society},
volume = {790},
number = {2},
pages = {87},
author = {Davis, Darren R. and Hayes, Wayne B.},
title = {SpArcFiRe: SCALABLE AUTOMATED DETECTION OF SPIRAL GALAXY ARM SEGMENTS},
journal = {The Astrophysical Journal}
}

@ARTICLE{DominguezSanchez+2019,
       author = {{Dom{\'\i}nguez S{\'a}nchez}, H. and {Huertas-Company}, M. and {Bernardi}, M. and {Kaviraj}, S. and {Fischer}, J.~L. and {Abbott}, T.~M.~C. and {Abdalla}, F.~B. and {Annis}, J. and {Avila}, S. and {Brooks}, D. and {Buckley-Geer}, E. and {Carnero Rosell}, A. and {Carrasco Kind}, M. and {Carretero}, J. and {Cunha}, C.~E. and {D'Andrea}, C.~B. and {da Costa}, L.~N. and {Davis}, C. and {De Vicente}, J. and {Doel}, P. and {Evrard}, A.~E. and {Fosalba}, P. and {Frieman}, J. and {Garc{\'\i}a-Bellido}, J. and {Gaztanaga}, E. and {Gerdes}, D.~W. and {Gruen}, D. and {Gruendl}, R.~A. and {Gschwend}, J. and {Gutierrez}, G. and {Hartley}, W.~G. and {Hollowood}, D.~L. and {Honscheid}, K. and {Hoyle}, B. and {James}, D.~J. and {Kuehn}, K. and {Kuropatkin}, N. and {Lahav}, O. and {Maia}, M.~A.~G. and {March}, M. and {Melchior}, P. and {Menanteau}, F. and {Miquel}, R. and {Nord}, B. and {Plazas}, A.~A. and {Sanchez}, E. and {Scarpine}, V. and {Schindler}, R. and {Schubnell}, M. and {Smith}, M. and {Smith}, R.~C. and {Soares-Santos}, M. and {Sobreira}, F. and {Suchyta}, E. and {Swanson}, M.~E.~C. and {Tarle}, G. and {Thomas}, D. and {Walker}, A.~R. and {Zuntz}, J.},
        title = "{Transfer learning for galaxy morphology from one survey to another}",
      journal = {\mnras},
         year = 2019,
        month = mar,
       volume = {484},
       number = {1},
        pages = {93-100},
          doi = {10.1093/mnras/sty3497},
archivePrefix = {arXiv},
       eprint = {1807.00807},
 primaryClass = {astro-ph.GA},
       adsurl = {https://ui.adsabs.harvard.edu/abs/2019MNRAS.484...93D}
}

@article{Ciprijanovic+2023,
doi = {10.1088/2632-2153/acca5f},
url = {https://dx.doi.org/10.1088/2632-2153/acca5f},
year = {2023},
month = {apr},
publisher = {IOP Publishing},
volume = {4},
number = {2},
pages = {025013},
author = {Ćiprijanović, A and Lewis, A and Pedro, K and Madireddy, S and Nord, B and Perdue, G N and Wild, S M},
title = {DeepAstroUDA: semi-supervised universal domain adaptation for cross-survey galaxy morphology classification and anomaly detection},
journal = {Machine Learning: Science and Technology}
}

@article{Cliff+1993,
author = {Cliff, Norman},
year = {1993},
month = {11},
pages = {494-509},
title = {Dominance Statistics: Ordinal Analyses to Answer Ordinal Questions},
volume = {114},
journal = {Psychological Bulletin},
doi = {10.1037/0033-2909.114.3.494}
}

@book{Cohen1988,
  author    = {Jacob Cohen},
  title     = {Statistical Power Analysis for the Behavioral Sciences},
  edition   = {2nd},
  year      = {1988},
  publisher = {Routledge},
  address   = {New York},
  doi       = {10.4324/9780203771587},
  url       = {https://doi.org/10.4324/9780203771587}
}

@ARTICLE{Siudek+2022,
       author = {{Siudek}, M. and {Ma{\l}ek}, K. and {Pollo}, A. and {Iovino}, A. and {Haines}, C.~P. and {Bolzonella}, M. and {Cucciati}, O. and {Gargiulo}, A. and {Granett}, B. and {Krywult}, J. and {Moutard}, T. and {Scodeggio}, M.},
        title = "{Shaping physical properties of galaxy subtypes in the VIPERS survey: Environment matters}",
      journal = {\aap},
         year = 2022,
        month = oct,
       volume = {666},
          eid = {A131},
        pages = {A131},
          doi = {10.1051/0004-6361/202243613},
archivePrefix = {arXiv},
       eprint = {2205.14736},
 primaryClass = {astro-ph.GA},
       adsurl = {https://ui.adsabs.harvard.edu/abs/2022A&A...666A.131S}
}

@ARTICLE{Huertas-Company+2015,
  title = {A {{CATALOG OF VISUAL-LIKE MORPHOLOGIES IN THE}} 5 {{CANDELS FIELDS USING DEEP LEARNING}}},
  author = {{Huertas-Company}, M. and Gravet, R. and {Cabrera-Vives}, G. and {P{\'e}rez-Gonz{\'a}lez}, P. G. and Kartaltepe, J. S. and Barro, G. and Bernardi, M. and Mei, S. and Shankar, F. and Dimauro, P. and Bell, E. F. and Kocevski, D. and Koo, D. C. and Faber, S. M. and Mcintosh, D. H.},
  year = {2015},
  journal = {The Astrophysical Journal Supplement Series},
  volume = {221},
  number = {1},
  pages = {8},
  doi = {10.1088/0067-0049/221/1/8},
  urldate = {2023-03-07},
  langid = {english}
}

@ARTICLE{Lotz+2008,
       author = {{Lotz}, Jennifer M. and {Davis}, M. and {Faber}, S.~M. and {Guhathakurta}, P. and {Gwyn}, S. and {Huang}, J. and {Koo}, D.~C. and {Le Floc'h}, E. and {Lin}, Lihwai and {Newman}, J. and {Noeske}, K. and {Papovich}, C. and {Willmer}, C.~N.~A. and {Coil}, A. and {Conselice}, C.~J. and {Cooper}, M. and {Hopkins}, A.~M. and {Metevier}, A. and {Primack}, J. and {Rieke}, G. and {Weiner}, B.~J.},
        title = "{The Evolution of Galaxy Mergers and Morphology at z < 1.2 in the Extended Groth Strip}",
      journal = {\apj},
         year = 2008,
        month = jan,
       volume = {672},
       number = {1},
        pages = {177-197},
          doi = {10.1086/523659},
archivePrefix = {arXiv},
       eprint = {astro-ph/0602088},
 primaryClass = {astro-ph},
       adsurl = {https://ui.adsabs.harvard.edu/abs/2008ApJ...672..177L}
}

@ARTICLE{Huertas-Company+2024,
       author = {{Huertas-Company}, M. and {Iyer}, K.~G. and {Angeloudi}, E. and {Bagley}, M.~B. and {Finkelstein}, S.~L. and {Kartaltepe}, J. and {McGrath}, E.~J. and {Sarmiento}, R. and {Vega-Ferrero}, J. and {Arrabal Haro}, P. and {Behroozi}, P. and {Buitrago}, F. and {Cheng}, Y. and {Costantin}, L. and {Dekel}, A. and {Dickinson}, M. and {Elbaz}, D. and {Grogin}, N.~A. and {Hathi}, N.~P. and {Holwerda}, B.~W. and {Koekemoer}, A.~M. and {Lucas}, R.~A. and {Papovich}, C. and {P{\'e}rez-Gonz{\'a}lez}, P.~G. and {Pirzkal}, N. and {Seill{\'e}}, L. -M. and {de la Vega}, A. and {Wuyts}, S. and {Yang}, G. and {Yung}, L.~Y.~A.},
        title = "{Galaxy morphology from z {\ensuremath{\sim}} 6 through the lens of JWST}",
      journal = {\aap},
         year = 2024,
        month = may,
       volume = {685},
          eid = {A48},
        pages = {A48},
          doi = {10.1051/0004-6361/202346800},
archivePrefix = {arXiv},
       eprint = {2305.02478},
 primaryClass = {astro-ph.GA},
       adsurl = {https://ui.adsabs.harvard.edu/abs/2024A&A...685A..48H}
}

@ARTICLE{Ghosh+2024,
       author = {{Ghosh}, Aritra and {Urry}, C. Megan and {Powell}, Meredith C. and {Shimakawa}, Rhythm and {van den Bosch}, Frank C. and {Nagai}, Daisuke and {Mitra}, Kaustav and {Connolly}, Andrew J.},
        title = "{Denser Environments Cultivate Larger Galaxies: A Comprehensive Study beyond the Local Universe with 3 Million Hyper Suprime-Cam Galaxies}",
      journal = {\apj},
         year = 2024,
        month = aug,
       volume = {971},
       number = {2},
          eid = {142},
        pages = {142},
          doi = {10.3847/1538-4357/ad596f},
archivePrefix = {arXiv},
       eprint = {2408.07128},
 primaryClass = {astro-ph.GA},
       adsurl = {https://ui.adsabs.harvard.edu/abs/2024ApJ...971..142G}
}

@article{Halkidi+2001,
  title={On Clustering Validation Techniques},
  author={Halkidi, Maria and Batistakis, Yannis and Vazirgiannis, Michalis},
  journal={Journal of Intelligent Information Systems},
  volume={17},
  number={2},
  pages={107--145},
  year={2001},
  month={dec},
  issn={1573-7675},
  doi={10.1023/A:1012801612483},
  url={https://doi.org/10.1023/A:1012801612483}
}

@article{Conselice+2000,
doi = {10.1086/308300},
url = {https://dx.doi.org/10.1086/308300},
year = {2000},
month = {feb},
publisher = {},
volume = {529},
number = {2},
pages = {886},
author = {Christopher J. Conselice and Matthew A. Bershady and Anna Jangren},
title = {The Asymmetry of Galaxies: Physical Morphology for Nearby and High-Redshift Galaxies},
journal = {The Astrophysical Journal}
}

@article{Gui+2024,
author = {Gui, Jie and Chen, Tuo and Zhang, Jing and Cao, Qiong and Sun, Zhenan and Luo, Hao and Tao, Dacheng},
title = {A Survey on Self-Supervised Learning: Algorithms, Applications, and Future Trends},
year = {2024},
issue_date = {Dec. 2024},
publisher = {IEEE Computer Society},
address = {USA},
volume = {46},
number = {12},
issn = {0162-8828},
url = {https://doi.org/10.1109/TPAMI.2024.3415112},
doi = {10.1109/TPAMI.2024.3415112},
journal = {IEEE Trans. Pattern Anal. Mach. Intell.},
month = dec,
pages = {9052–9071},
numpages = {20}
}

@article{Russakovsky+2015,
author = {Russakovsky, Olga and Deng, Jia and Su, Hao and Krause, Jonathan and Satheesh, Sanjeev and Ma, Sean and Huang, Zhiheng and Karpathy, Andrej and Khosla, Aditya and Bernstein, Michael and Berg, Alexander C. and Fei-Fei, Li},
title = {ImageNet Large Scale Visual Recognition Challenge},
year = {2015},
issue_date = {December 2015},
publisher = {Kluwer Academic Publishers},
address = {USA},
volume = {115},
number = {3},
issn = {0920-5691},
url = {https://doi.org/10.1007/s11263-015-0816-y},
doi = {10.1007/s11263-015-0816-y},
journal = {Int. J. Comput. Vision},
month = dec,
pages = {211–252},
numpages = {42}
}

@ARTICLE{ge+2023,
       author = {{Zhu-Ge}, Jia-Ming and {Luo}, Jia-Wei and {Zhang}, Bing},
        title = "{Machine learning classification of CHIME fast radio bursts - II. Unsupervised methods}",
      journal = {\mnras},
         year = 2023,
        month = feb,
       volume = {519},
       number = {2},
        pages = {1823-1836},
          doi = {10.1093/mnras/stac3599},
archivePrefix = {arXiv},
       eprint = {2210.02471},
 primaryClass = {astro-ph.HE},
       adsurl = {https://ui.adsabs.harvard.edu/abs/2023MNRAS.519.1823Z}
}

@ARTICLE{Davies+1979,
  author={Davies, David L. and Bouldin, Donald W.},
  journal={IEEE Transactions on Pattern Analysis and Machine Intelligence}, 
  title={A Cluster Separation Measure}, 
  year={1979},
  volume={PAMI-1},
  number={2},
  pages={224-227},
  doi={10.1109/TPAMI.1979.4766909}}

@article{Freeman+2013,
    author = {Freeman, P. E. and Izbicki, R. and Lee, A. B. and Newman, J. A. and Conselice, C. J. and Koekemoer, A. M. and Lotz, J. M. and Mozena, M.},
    title = "{New image statistics for detecting disturbed galaxy morphologies at high redshift}",
    journal = {Monthly Notices of the Royal Astronomical Society},
    volume = {434},
    number = {1},
    pages = {282-295},
    year = {2013},
    month = {06},
    issn = {0035-8711},
    doi = {10.1093/mnras/stt1016},
    url = {https://doi.org/10.1093/mnras/stt1016},
    eprint = {https://academic.oup.com/mnras/article-pdf/434/1/282/18499355/stt1016.pdf},
}

@article{Lotz+2006,
doi = {10.1086/497950},
url = {https://dx.doi.org/10.1086/497950},
year = {2006},
month = {jan},
publisher = {},
volume = {636},
number = {2},
pages = {592},
author = {Jennifer M. Lotz and Piero Madau and Mauro Giavalisco and Joel Primack and Henry C. Ferguson},
title = {The Rest-Frame Far-Ultraviolet Morphologies of Star-Forming Galaxies at z ~ 1.5 and 4},
journal = {The Astrophysical Journal}
}

@InProceedings{Masci+2011,
author="Masci, Jonathan
and Meier, Ueli
and Cire{\c{s}}an, Dan
and Schmidhuber, J{\"u}rgen",
editor="Honkela, Timo
and Duch, W{\l}odzis{\l}aw
and Girolami, Mark
and Kaski, Samuel",
title="Stacked Convolutional Auto-Encoders for Hierarchical Feature Extraction",
booktitle="Artificial Neural Networks and Machine Learning -- ICANN 2011",
year="2011",
publisher="Springer Berlin Heidelberg",
address="Berlin, Heidelberg",
pages="52--59",
isbn="978-3-642-21735-7"
}

@article{Dai+2023,
doi = {10.3847/1538-4365/ace69e},
url = {https://dx.doi.org/10.3847/1538-4365/ace69e},
year = {2023},
month = {sep},
publisher = {The American Astronomical Society},
volume = {268},
number = {1},
pages = {34},
author = {Yao Dai and Jun Xu and Jie Song and Guanwen Fang and Chichun Zhou and Shuo Ba and Yizhou Gu and Zesen Lin and Xu Kong},
title = {The Classification of Galaxy Morphology in the H Band of the COSMOS-DASH Field: A Combination-based Machine-learning Clustering Model},
journal = {The Astrophysical Journal Supplement Series}
}

@article{Andrzej+1993,
title = {Principal components analysis (PCA)},
journal = {Computers \& Geosciences},
volume = {19},
number = {3},
pages = {303-342},
year = {1993},
issn = {0098-3004},
doi = {https://doi.org/10.1016/0098-3004(93)90090-R},
url = {https://www.sciencedirect.com/science/article/pii/009830049390090R},
author = {Andrzej Maćkiewicz and Waldemar Ratajczak}
}

@article{Zhang+1996,
author = {Zhang, Tian and Ramakrishnan, Raghu and Livny, Miron},
title = {BIRCH: an efficient data clustering method for very large databases},
year = {1996},
issue_date = {June 1996},
publisher = {Association for Computing Machinery},
address = {New York, NY, USA},
volume = {25},
number = {2},
issn = {0163-5808},
url = {https://doi.org/10.1145/235968.233324},
doi = {10.1145/235968.233324},
journal = {SIGMOD Rec.},
month = {jun},
pages = {103–114},
numpages = {12}
}

@article{Cheng2+2021,
    author = {Cheng, Ting-Yun and Huertas-Company, Marc and Conselice, Christopher J and Aragón-Salamanca, Alfonso and Robertson, Brant E and Ramachandra, Nesar},
    title = {Beyond the hubble sequence – exploring galaxy morphology with unsupervised machine learning},
    journal = {Monthly Notices of the Royal Astronomical Society},
    volume = {503},
    number = {3},
    pages = {4446-4465},
    year = {2021},
    month = {03},
    issn = {0035-8711},
    doi = {10.1093/mnras/stab734},
    url = {https://doi.org/10.1093/mnras/stab734},
    eprint = {https://academic.oup.com/mnras/article-pdf/503/3/4446/39066819/stab734.pdf},
}

@ARTICLE{Willett+2017,
       author = {{Willett}, Kyle W. and {Galloway}, Melanie A. and {Bamford}, Steven P. and {Lintott}, Chris J. and {Masters}, Karen L. and {Scarlata}, Claudia and {Simmons}, B.~D. and {Beck}, Melanie and {Cardamone}, Carolin N. and {Cheung}, Edmond and {Edmondson}, Edward M. and {Fortson}, Lucy F. and {Griffith}, Roger L. and {H{\"a}u{\ss}ler}, Boris and {Han}, Anna and {Hart}, Ross and {Melvin}, Thomas and {Parrish}, Michael and {Schawinski}, Kevin and {Smethurst}, R.~J. and {Smith}, Arfon M.},
        title = "{Galaxy Zoo: morphological classifications for 120 000 galaxies in HST legacy imaging}",
      journal = {\mnras},
         year = 2017,
        month = feb,
       volume = {464},
       number = {4},
        pages = {4176-4203},
          doi = {10.1093/mnras/stw2568},
archivePrefix = {arXiv},
       eprint = {1610.03068},
 primaryClass = {astro-ph.GA},
       adsurl = {https://ui.adsabs.harvard.edu/abs/2017MNRAS.464.4176W}
}

@article{Kolmogorov+1933,
  title   = {Sulla determinazione empirica di una legge di distribuzione},
  author  = {Kolmogorov, Andrej Nikolajevi{\v{c}}},  
  journal = {Giornale dell'Istituto Italiano degli Attuari},
  year    = {1933},
  volume  = {4},
  pages   = {83--91},
  url     = {https://api.semanticscholar.org/CorpusID:222427298}
}

@article{Murtagh+1983,
    author = {Murtagh, F.},
    title = "{A Survey of Recent Advances in Hierarchical Clustering Algorithms}",
    journal = {The Computer Journal},
    volume = {26},
    number = {4},
    pages = {354-359},
    year = {1983},
    month = {11},
    issn = {0010-4620},
    doi = {10.1093/comjnl/26.4.354},
    url = {https://doi.org/10.1093/comjnl/26.4.354},
    eprint = {https://academic.oup.com/comjnl/article-pdf/26/4/354/1072603/26-4-354.pdf},
}

@article{Murtagh+2014,
   title={Ward’s Hierarchical Agglomerative Clustering Method: Which Algorithms Implement Ward’s Criterion?},
   volume={31},
   ISSN={1432-1343},
   url={http://dx.doi.org/10.1007/s00357-014-9161-z},
   DOI={10.1007/s00357-014-9161-z},
   number={3},
   journal={Journal of Classification},
   publisher={Springer Science and Business Media LLC},
   author={Murtagh, Fionn and Legendre, Pierre},
   year={2014},
   month=oct, pages={274–295} }

@article{Hartigan+1979,
 ISSN = {00359254, 14679876},
 URL = {http://www.jstor.org/stable/2346830},
 author = {J. A. Hartigan and M. A. Wong},
 journal = {Journal of the Royal Statistical Society. Series C (Applied Statistics)},
 number = {1},
 pages = {100--108},
 publisher = {[Wiley, Royal Statistical Society]},
 title = {Algorithm AS 136: A K-Means Clustering Algorithm},
 urldate = {2024-03-13},
 volume = {28},
 year = {1979}
}

@misc{liu+2023,
  author  = {Liu, Zhaocong and Zhang, Fa and Cheng, Lin and Deng, Huanxi and Yang, Xiaoyan and Zhang, Zhenyu and Zhou, Chi-Chun},
  title   = {Simple But Effective Unsupervised Classification for Specified Domain Images: A Case Study on Fungi Images},
  year    = {2023},
  howpublished = {\url{https://ssrn.com/abstract=4673082}},
  doi     = {10.2139/ssrn.4673082}
}

@article{Krizhevsky+2012,
  title={ImageNet classification with deep convolutional neural networks},
  author={Alex Krizhevsky and Ilya Sutskever and Geoffrey E. Hinton},
  journal={Communications of the ACM},
  year={2012},
  volume={60},
  pages={84 - 90},
  url={https://api.semanticscholar.org/CorpusID:195908774}
}

@article{Yao+2023,
doi = {10.3847/1538-4357/ace7b5},
url = {https://dx.doi.org/10.3847/1538-4357/ace7b5},
year = {2023},
month = {aug},
publisher = {The American Astronomical Society},
volume = {954},
number = {2},
pages = {113},
author = {Yao Yao and Jie Song and Xu Kong and Guanwen Fang and Hong-Xin Zhang and Xinkai Chen},
title = {Evolution of Nonparametric Morphology of Galaxies in the JWST CEERS Field at z ≃ 0.8–3.0},
journal = {The Astrophysical Journal}
}

@article{Barden+2012,
  title={GALAPAGOS: From Pixels to Parameters},
  author={Marco Barden and Boris Haussler and Chien Y. Peng and D. H. Mcintosh and Yicheng Guo},
  journal={Monthly Notices of the Royal Astronomical Society},
  year={2012},
  volume={422},
  pages={449-468},
  url={https://api.semanticscholar.org/CorpusID:119117461}
}

@article{Koekemoer+2007,
doi = {10.1086/520086},
url = {https://dx.doi.org/10.1086/520086},
year = {2007},
month = {sep},
publisher = {},
volume = {172},
number = {1},
pages = {196},
author = {A. M. Koekemoer and H. Aussel and D. Calzetti and P. Capak and M. Giavalisco and J.-P. Kneib and A. Leauthaud and O. Le Fèvre and H. J. McCracken and R. Massey and B. Mobasher and J. Rhodes and N. Scoville and P. L. Shopbell},
title = {The COSMOS Survey: Hubble Space Telescope Advanced Camera for Surveys Observations and Data Processing*},
journal = {The Astrophysical Journal Supplement Series}
}

@INPROCEEDINGS{Koekemoer+2003,
       author = {{Koekemoer}, A.~M. and {Fruchter}, A.~S. and {Hook}, R.~N. and {Hack}, W.},
        title = "{MultiDrizzle: An Integrated Pyraf Script for Registering, Cleaning and Combining Images}",
    booktitle = {HST Calibration Workshop : Hubble after the Installation of the ACS and the NICMOS Cooling System},
         year = 2003,
        month = jan,
        pages = {337},
       adsurl = {https://ui.adsabs.harvard.edu/abs/2003hstc.conf..337K}
}

@article{Song+2024,
doi = {10.3847/1538-4365/ad434f},
url = {https://dx.doi.org/10.3847/1538-4365/ad434f},
year = {2024},
month = {jun},
publisher = {The American Astronomical Society},
volume = {272},
number = {2},
pages = {42},
author = {Jie Song and GuanWen Fang and Shuo Ba and Zesen Lin and Yizhou Gu and Chichun Zhou and Tao Wang and Cai-Na Hao and Guilin Liu and Hongxin Zhang and Yao Yao and Xu Kong},
title = {USmorph: An Updated Framework of Automatic Classification of Galaxy Morphologies and Its Application to Galaxies in the COSMOS Field},
journal = {The Astrophysical Journal Supplement Series}
}

@ARTICLE{haubler+2022,
       author = {{H{\"a}u{\ss}ler}, Boris and {Vika}, Marina and {Bamford}, Steven P. and {Johnston}, Evelyn J. and {Brough}, Sarah and {Casura}, Sarah and {Holwerda}, Benne W. and {Kelvin}, Lee S. and {Popescu}, Cristina},
        title = "{GALAPAGOS-2/GALFITM/GAMA - Multi-wavelength measurement of galaxy structure: Separating the properties of spheroid and disk components in modern surveys}",
      journal = {\aap},
         year = 2022,
        month = aug,
       volume = {664},
          eid = {A92},
        pages = {A92},
          doi = {10.1051/0004-6361/202142935},
archivePrefix = {arXiv},
       eprint = {2204.05907},
 primaryClass = {astro-ph.GA},
       adsurl = {https://ui.adsabs.harvard.edu/abs/2022A&A...664A..92H}
}

\end{document}